\documentclass[12pt]{article}
\usepackage{lineno}
\usepackage{newtxtext,newtxmath}

\usepackage{graphicx}

\usepackage[letterpaper,margin=1in]{geometry}

\renewenvironment{abstract}
	{\quotation}
	{\endquotation}

\date{}

\makeatletter
\renewcommand{\fnum@figure}{\textbf{Figure \thefigure}}
\renewcommand{\fnum@table}{\textbf{Table \thetable}}
\makeatother

\usepackage{scicite}

\usepackage{url}

\newcommand{\add}{ }
\newcommand{\addmore}{}

\def\scititle{
	Large-amplitude Variability Driven by Giant Dust Storms on a Planetary-mass Companion
}
\title{\bfseries \boldmath \scititle}

\author{
	Xianyu Tan$^{1\ast}$,
	Xi Zhang$^{2}$,
	Mark S. Marley$^{3}$,
        Yifan Zhou$^{4}$,
        Ben W. P. Lew$^{5}$,\and
        Brittany E. Miles$^{6}$,
        Natasha E. Batalha$^{7}$,
        Beth A. Biller$^{8,9}$,
        Gaël Chauvin$^{10}$,\and
        Sasha Hinkley$^{11}$,
        Kielan K. W. Hoch$^{12}$,
        Elena Manjavacas$^{13}$,
        Stanimir Metchev$^{14}$,\and
        Simon Petrus$^{15}$,
        Emily Rickman$^{13}$,
        Andrew Skemer$^{16}$,
        Genaro Suárez$^{17}$,\and
        Ben J. Sutlieff$^{8,9}$,
        Johanna M. Vos$^{18}$,
        Niall Whiteford$^{17}$\and
        \footnotesize$^{1}$Tsung-Dao Lee Institute \& School of Physics and Astronomy, Shanghai Jiao Tong University, Shanghai 201210, China\and
	\footnotesize$^{2}$Department of Earth and Planetary Sciences, University of California Santa Cruz, Santa Cruz, CA 95064, USA\and
        \footnotesize$^{3}$Lunar and Planetary Laboratory, The University of Arizona, 1629 E. University Blvd, Tucson, AZ 85721, USA\and
        \footnotesize$^{4}$Department of Astronomy, University of Virginia, 530 McCormick Rd, Charlottesville, VA 22904, USA\and
        \footnotesize$^{5}$Bay Area Environmental Research Institute and NASA Ames Research Center, Moffett Field, CA 94035, USA\and
        \footnotesize$^{6}$Department of Astronomy and Steward Observatory, The University of Arizona, 933 N Cherry Ave., Tucson, AZ 85721, USA\and
        \footnotesize$^{7}$NASA Ames Research Center, Moffett Field, CA 94035, USA\and
        \footnotesize$^{8}$Scottish Universities Physics Alliance, Institute for Astronomy, University of Edinburgh, \and
        \footnotesize Blackford Hill, Edinburgh EH9 3HJ, UK\and
        \footnotesize$^{9}$Centre for Exoplanet Science, University of Edinburgh, Edinburgh EH9 3HJ, UK\and
        \footnotesize$^{10}$Max Planck Institute for Astronomy, Koenigstuhl 17, D-69117 Heidelberg, Germany\and
        \footnotesize$^{11}$University of Exeter, Astrophysics Group, Physics Building, Stocker Road, Exeter, EX4 4QL, UK\and
        \footnotesize$^{12}$Space Telescope Science Institute, Baltimore, MD 21218, USA\and
        \footnotesize$^{13}$European Space Agency (ESA), ESA Office, Space Telescope Science Institute, \and
        \footnotesize 3700 San Martin Drive, Baltimore, MD, 21218 USA\and
        \footnotesize$^{14}$Western University, Department of Physics \& Astronomy and Institute for Earth and Space Exploration, \and
        \footnotesize 1151 Richmond Street, London, Ontario N6A 3K7, Canada\and
        \footnotesize$^{15}$NASA Goddard Space Flight Center, Greenbelt, MD 20771, USA\and
        \footnotesize$^{16}$Astronomy \& Astrophysics Department, University of California Santa Cruz, Santa Cruz, CA 95064, USA\and
        \footnotesize$^{17}$Department of Astrophysics, American Museum of Natural History,\and 
        \footnotesize Central Park West at 79th Street, New York, NY 10034, USA\and
        \footnotesize$^{18}$School of Physics, Trinity College Dublin, The University of Dublin, Dublin 2, Ireland\and
	\small$^\ast$Corresponding author. Email: \url{xianyut@sjtu.edu.cn}
}

\begin{document} 

\maketitle

\begin{abstract} \bfseries \boldmath
Large-amplitude variations are commonly observed in the atmospheres of directly imaged exoplanets and brown dwarfs. VHS 1256B, the most variable known planet-mass object, exhibits a near-infrared flux change of nearly 40\%, with red color and silicate features revealed in recent JWST spectra, challenging current theories. Using a general circulation model, we demonstrate that VHS 1256B’s atmosphere is dominated by planetary-scale dust storms persisting for tens of days, with large patchy clouds propagating with equatorial waves. This weather pattern, distinct from the banded structures seen on solar system giants, simultaneously explains the observed spectra and critical features in the rotational light curves, including the large amplitude, irregular evolution, and wavelength dependence, as well as the variability trends observed in near-infrared color-magnitude diagrams of {\add dusty} substellar atmospheres.

\end{abstract}

\section{Introduction}
Directly imaged planets and brown dwarfs are extra-solar analogs of solar system giant planets, with most detected to date being high-temperature, massive counterparts that share similarities in size, rotation, and atmospheric constituents \cite{burrows2001}. However, resolving the surface maps of these distant bodies remains challenging. Rotational variability serves as a valuable tool for constraining their weather patterns, as observed variations are driven by inhomogeneities in temperature, composition, and cloud distribution \cite{karalidi2015,biller2017, apai2017, artigau2018,ge2019rotational}.

Rotational modulations on solar system giants like Jupiter and Neptune have been investigated. Jupiter’s variability is dominated by discrete atmospheric features such as the Great Red Spot and hot spots in the North Equatorial Belt \cite{ge2019rotational,marley1999giant}. On Neptune, temperature anomalies and patchy clouds in different zones cause the rotational modulations observed in scattered and thermal light \cite{simon2016,Stauffer2016, rowe2017, apai2017}. Typically, rotational modulation on giant planets results in a few percent variation in reflection and emission. By analogy, it was proposed that spots and {\add wave beating patterns} in the zonal (east-west) bands could explain similar light curve variability of some brown dwarfs with amplitudes also around a few percent \cite{karalidi2015, apai2017,apai2021,fuda2024,fuda2024polar}. {\add However, unlike on giant planets where moist convection and differential solar heating are key drivers of atmospheric dynamics \cite{lian&showman2010,liu2011}, these processes are not significant on hot brown dwarfs \cite{tan2017}. Therefore, the physical origins of spots and differentially propagating waves on these objects remain elusive.}

{\add Moreover}, a fraction of these substellar bodies exhibit exceptionally large variation amplitudes (e.g., \cite{radigan2012, apai2013, buenzli2015b, apai2017, vos2019, zhou2022}). Among them, VHS J125601.92–125723.9 B (hereafter VHS 1256B) stands out as the most variable known to date, with a near-infrared flux change of nearly 40\% (\cite{bowler2020, zhou2020, zhou2022, miles2023}, see also fig. \ref{fig.obs_lc}). VHS 1256B is an L-to-T transition object with a mass of less than 20 Jupiter masses \cite{dupuy2023,miles2023} and a rotation period of approximately 22 hours \cite{zhou2020}. In the solar system, only persistent features like Jupiter’s Great Red Spot and hot spots can produce variations of around 20\% in the mid-infrared \cite{ge2019rotational}, yet VHS 1256B also shows substantial inter-annual variability, in stark contrast to the long-term stability of the Great Red Spot. 
{\add In addition}, VHS 1256B’s variation amplitude varies significantly with wavelength, from a few tens of percent in the near-infrared to a few percent in the mid-infrared \cite{zhou2020, zhou2022}, a pattern distinct from Jupiter’s modulation \cite{ge2019rotational}.

Recent JWST observations of VHS 1256B have revealed a strikingly red spectrum with prominent silicate features at 10 microns \cite{miles2023}, indicating that silicate clouds likely play a critical role in driving its atmospheric dynamics and observed characteristics. VHS 1256B’s extreme, evolving light-curve variability, combined with its distinct red and dusty spectrum, positions it as a vital target for redefining theories about weather processes in substellar atmospheres. The physical mechanisms behind its variability, cloud formation, and atmospheric evolution remain poorly understood. To date, no model has successfully explained both the spectra and wavelength-dependent light curves of directly imaged exoplanets in a self-consistent manner.

\section{Results}
\subsection*{Giant dust storm formation}
Here we demonstrate that VHS 1256B’s atmospheric weather is primarily driven by massive, recurring dust storms manifested as global-scale tropical waves, which effectively explain both its time-resolved spectral and photometric characteristics. These dust storms, composed of silicate {\add and iron} clouds, shape the planet’s red spectrum and are responsible for the observed variations in its light curves across different wavelengths and epochs. We simulated the three-dimensional (3D) atmospheric dynamics of VHS 1256B using a general circulation model (GCM) tailored for substellar atmospheres \cite{tan2019, tan2021bd1, tan2021bd2}, incorporating cloud formation and equilibrium chemistry processes \cite{ackerman2001, visscher2010} (see Supplementary Materials). We found that cloud radiative feedback is the primary driver of these atmospheres' circulation and surface inhomogeneity. When clouds form, they exert greenhouse warming on the cloudy layers, whereas in relatively cloud-free regions flux can escape from deeper layers and cool the column. The horizontal heating and cooling contrast drives a large-scale circulation that maintains and triggers cloud cycles. Due to the planetary rotation, storms typically organize themselves as vortices at high latitudes with chaotic evolution of individual vortices  \cite{tan2021bd1}. 

\begin{figure}
    \centering
    \includegraphics[width=1\textwidth]{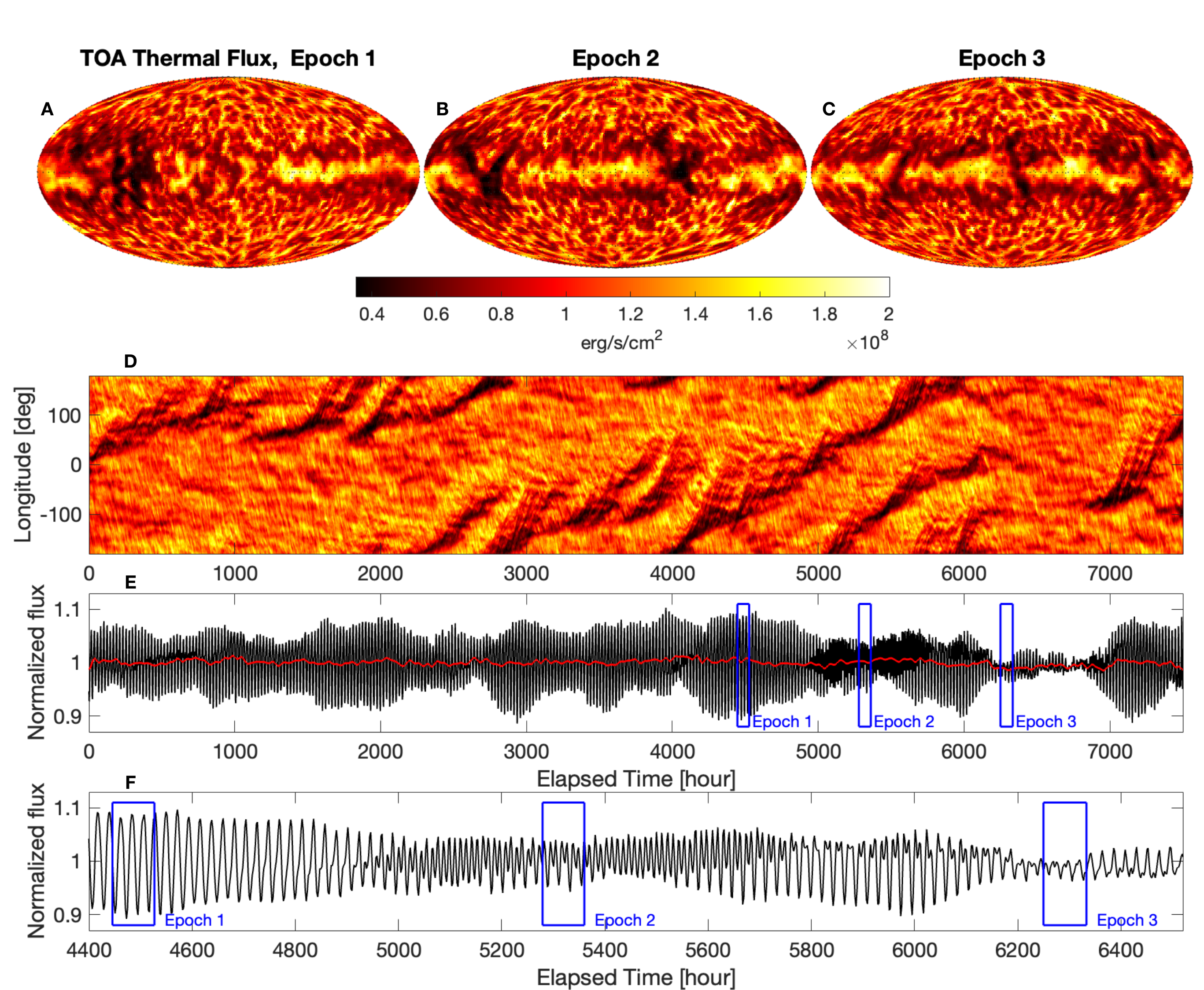}
    \caption{\textbf{Bolometric thermal flux maps, their time evolution near the equator, and the resulting long-term light-curve evolution from the GCM.}
    {\add \textit{Panel A, B, C:} Top-of-atmosphere (TOA) bolometric flux maps from the GCM at different epochs, showing global-scale dust storms (the dark regions) at the equator. } \textit{Panel D:} the Hovmoller diagram (longitude-time sequence) of the TOA bolometric flux near the equator. The flux is averaged over latitudes within $\pm 10$ degrees to cover a fraction of the meridionally extended waves. The dark features evolving towards the upper right mean the eastward propagation of the dust storms.  \textit{Panel E:} the long-term white light curve of the GCM (black line). The red line is the light curve with periodic signals less than 4 rotation periods filtered out. The much smaller amplitude of the filtered light curve suggests that the variability (black line) is mostly rotational rather than from an intrinsic variation of the mean atmospheric state. Epochs 1, 2, and 3 enclosed in the panel correspond to {\add a few different characteristic states of the equatorial waves}. {\add \textit{Panel F:} a zoom-in of the light curve between elapsed time of 4400 to 6500 hours.}
    }
    \label{fig.whitelight}
\end{figure}

However, {\add at low latitudes,  the meridional (north-south) gradient of the Coriolis parameter, $\beta=df/dy$ where $f$ is the Coriolis parameter, sculpts the large-scale flow and forms tropical waves. Rossby waves are triggered off the equator and initially travel west of the cloudy region; Kelvin waves form at the equator and initially travel east of the cloudy region. They are then sustained by cloud radiative heating near the equator and reach a slowly evolving ``Chevron" pattern, forming massive storms that propagate zonally \cite{tan2021bd2}.  
Instantaneous global thermal flux maps at different epochs of the model outputs are shown in panel A, B, C of Fig. \ref{fig.whitelight},} in which the dark region at the equator corresponds to a vigorously upwelling cloud forming zone {\add east} of the equatorial Kelvin wave and a long strip of bright region westward of the dark region corresponds to a subsiding, relatively clear-sky zone of the wave. Off-equatorial dark strips aligned with the bright equatorial strip are associated with the {\add cloudy areas equatorward of the Rossby waves. }


The characteristic size of the tropical storms depends on the rotation of the planet. Their meridional extent roughly scales with the equatorial deformation radius $L_{\beta} =\sqrt{c/\beta}$, where $c$ is the horizontal phase speed of gravity waves \cite{tan2021bd2}. For a slow rotator like VHS 1256B, $\beta$ is relatively small, and the half meridional width of the tropical storm is roughly 30,000 km, nearly half of the planetary radius. In the zonal direction, most of the time, a single tropical storm {\add is the most dominant feature over} a full zonal circle. This zonal scale is related to the overturning circulation associated with cloud feedback, which tends to be on the planetary scale at the temperature range similar to VHS 1256B \cite{tan2021bd1}. As the body rotates, the giant tropical storm appears in and out of view, resulting in a significant light curve variability as shown in panel F of Fig. \ref{fig.whitelight}. The wavelength-integrated bolometric light curve shows nearly 20\% peak-to-peak amplitude at its most variable epochs (e.g., epoch 1), with the maximum variation of nearly 40\% in the near-infrared wavelengths, consistent with that observed by the Hubble Space Telescope (\cite{zhou2022}, see also the observed light curves in fig. \ref{fig.obs_lc}).

\begin{figure}
    \centering
    \includegraphics[width=0.6\linewidth]{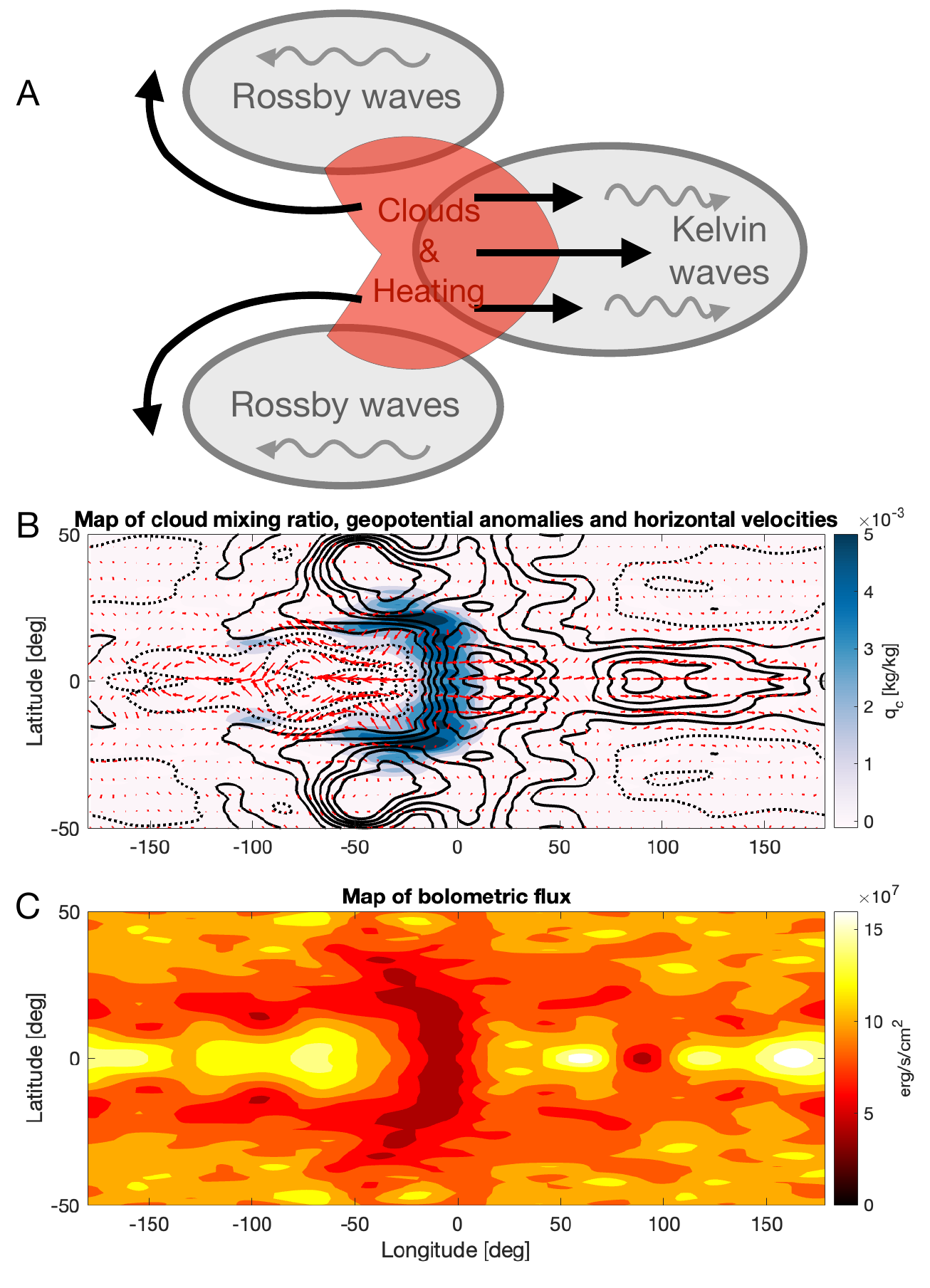}
    \caption{{\add \textbf{A sketch of the wave generation mechanism and wave diagnoses based on GCM outputs.} {\it Panel A}: schematic of the horizontal structure of idealized tropical waves relevant to our GCM results. Red shading shows the cloud formation and radiatively heated region at the equator; initially westward-propagating Rossby waves (with wiggle arrows representing propagation direction) are off-equator, and initially eastward-propagating Kelvin waves are east of the heating region. Wave velocities are shown as black arrows. The sketch is similar to that shown in Vallis 2021 \cite{vallis2021}. {\it Panel B:} diagnosis of cloud structure (color), geopotential anomalies (solid lines representing positive values and dotted lines representing negative values), and horizontal velocity fields (red arrows). {\it Panel C:} top-of-atmosphere thermal flux corresponding to the cloud structure shown in panel B. }
    }\label{fig.mechanism}
\end{figure}

{\add  A schematic is shown in Fig. \ref{fig.mechanism} to illustrate the tropical wave dynamical mechanism in our GCM. The pioneer work by Matsuno in 1966 \cite{matsuno1966} revealed that the large-scale atmospheric disturbances take the form of equatorially trapped free waves, including westward-propagating Rossby waves and eastward-propagating Kelvin waves. When these waves are sustained by a stationary heating source centered at the equator and balanced by frictional drag, standing wave solutions were obtained by Gill in 1980 \cite{gill1980}, characterized by two off-equatorial Rossby waves held west of the heating source and a Kelvin wave held east of the heating source. This so-called Matsuno-Gill wave pattern has far-reaching applications to a variety of atmospheric phenomena ranging from tropical variabilities on Earth \cite{zhang2020MJOreview}, to the circulation of close-in exoplanets \cite{showman2011}, and to the variability of isolated giant planets and brown dwarfs \cite{tan2021bd2}.}

{\add In our models, when clouds form by convergence at the equator, their radiative heating triggers the tropical waves, and the quasi-Matsuno-Gill shape forms around the cloudy region over wave-propagation timescales. However, clouds are at the same time advected by the flows over a timescale comparable to the wave propagation timescale. Unlike the classical Matsuno-Gill standing waves, the heat source is nonstationary and is coupled with the waves. Such a weakly nonlinear coupling leads to the slow propagation of the entire wave pattern (as shown in panel D of Fig. \ref{fig.whitelight}) and rich dynamical phenomena, which are still poorly understood \cite{vallis2021,tan2021bd2}. Panel A in Fig. \ref{fig.mechanism} {\addmore displays} a schematic that depicts the above picture: cloud heating at the equator initially triggers Rossby and Kelvin waves, and the cloud distribution is subsequently shaped by the waves. Panel B shows a GCM diagnosis of cloud mixing ratio, geopotential anomalies and eddy velocities associated with these waves (for technical details, readers are referred to Tan \& Showman 2021 \cite{tan2021bd2}). Panel C shows the corresponding top-of-atmosphere thermal flux, quantifying the relation between the tropical waves, thermal flux inhomogeneity, and the resulting light-curve variability. As shown in panel B, clouds are cleared out west of the quasi-standing Kelvin wave near the cloud-forming region due to subsiding motions. There are likely more complications contributing to this cloud-clearing process in addition to the quasi-stationary waves near the cloud-forming region. For example, as shown in panel B, extra traveling waves extend far outside the cloudy region and merge into the east edge of the cloudy region, which may generate a subsiding region favorable for cloud clearing. These subtleties may be cleanly explained only by a more dedicated dynamical theory in future work. For more discussion of tropical wave dynamics in a related context, readers are referred to Vallis 2021 \cite{vallis2021}. }

\subsection*{Irregular light curve evolution}
The simulated tropical wave shows significant evolution over timescales spanning from tens to hundreds of days, much longer than the rotation period. {\add Maps at different epochs are shown in panels A, B, and C, and the tropical time evolution over 300 rotation periods comprising of both zonal propagations and the dissipation and resurgence of tropical storms is shown in panel D of Fig. \ref{fig.whitelight}.} The time evolution results in a long-term irregularity of the bolometric light curve shown in panel E of Fig. \ref{fig.whitelight}. The irregular evolution is not due to the intrinsic oscillation of the mean atmospheric state but originates from the changing morphology of the storms.  Most of the time, the equatorial storm comprises {\add mainly} a single wavelet with a zonal wavenumber=1. This maximizes the light-curve variability and signals a single sinusoid over a rotation period {\add (for example, the light curve at epoch 1 as zoomed in in panel F of Fig. \ref{fig.whitelight}).} {\add There are times when a new tropical storm emerges and two tropical waves coexist at the equator, taking a zonal wavenumber=2 feature (epoch 2 in panel B of Fig. \ref{fig.whitelight})} which generates a double-peak light curve over one rotation period such as that shown near epoch 2. There are also times when {\add the atmosphere does not host a single or two major tropical storms (epoch 3 in panel C of Fig. \ref{fig.whitelight}),} substantially reducing the cloud contrast and the light-curve amplitude (see the evolution near epoch 3). Certain irregularities appear over a rotational timescale near epoch 3 because subtropical fast-evolving vortices also contribute to the light curve. 

The nonlinear tropical storm evolution provides a dynamical mechanism to explain why the second set of HST measurements taken in 2020 revealed less variability and certain nonperiodic variations (\cite{zhou2022}, and fig. \ref{fig.obs_lc}). The irregular evolution of light curves has been commonly observed in many isolated brown dwarfs and planetary-mass objects, although typically with smaller amplitudes (e.g., \cite{artigau2009,metchev2015,apai2017,apai2021,fuda2024}). It has been suggested that differential propagating waves in the zonal bands could produce beating patterns such as that on Neptune, potentially explaining the variability in light curves \cite{apai2017,apai2021}. However, our simulations of VHS 1256B reveal a distinctly different physical mechanism from the banded giant planets in the solar system. The simulated brightness map does not exhibit a banded structure (Fig. \ref{fig.whitelight} and fig. \ref{fig.maps}) but is significantly regulated by the patchy dusty clouds in the evolving tropical wave system, which naturally explains the puzzling variability observed in the light curves.

{\add This distinction from Solar System planets arises because the atmospheric circulation regime of hot brown dwarfs is fundamentally different. One of the key physical parameters is the ratio of the radiative timescale to the dynamical timescale. In Jupiter and Saturn, this ratio is very large (long radiative times), whereas for hot brown dwarfs, it is of order unity or smaller \cite{showman2020,zhang2020}. The atmospheres of Jupiter and Saturn are dominated by zonal jets that occupy the majority of the total kinetic energy and whose meridional widths are a small fraction of the planetary radius. Within the jets, eddies (waves, turbulence, and vortices) are generated on much smaller length scales and grow in size via strong nonlinear interactions \cite{ingersoll2004}. Decades of observational, theoretical and numerical studies of their circulation suggest a basic framework that the zonal jets are pumped up by small-scale eddies via a ``slow" upscale energy transport, despite various eddy generation and energy transport mechanisms being proposed \cite{vasavada2005,showman2018review}. Furthermore, the roles of clouds on the jet generation and dynamics on Jupiter and Saturn remain elusive \cite{fletcher2020well}. Zonally differential propagation of discrete features such as the Great Red Spot embedded in the alternating jets similar to Jupiter and Saturn has been proposed as one of the phenomena related to the complex light-curve behaviors of brown dwarfs \cite{apai2021,fuda2024}, which implies that the underlying dynamical mechanism of these light curves may involve strongly nonlinear, progressive upscale energy transfer.} 

{\add In contrast, the light-curve variability shown in our GCMs is mainly tied to large-scale tropical waves whose formation process is quasi-linear, where the cloud radiative feedback is essential for the wave generation. The waves obtain their energy via a direct wave adjustment process and grow to a size comparable to a planetary radius over a short timescale \cite{tan2021bd1}. Arguably, the extremely efficient radiative heating or cooling in the short-radiative-timescale regime for observed brown dwarfs is the key for generating such circulation patterns with significant spatial contrasts of cloud and temperature structures \cite{showman2020}, whereas it could serve as an obstacle for upscale energy pumping mechanisms for the Jovian-like regime \cite{zhang2014,showman2019}. In addition, some processes, including moist convection and baroclinic instability associated with solar differential insolation \cite{lian&showman2010,schneider2009} that are thought to play critical roles in pumping the circulation of Jupiter and Saturn, be negligible in the regime of brown dwarf atmospheres \cite{tan2017}.  Therefore, the modeled light-curve variability in this work does not involve the ``slow" jet-forming mechanisms similar to those of Jupiter and Saturn.  }

\begin{figure}
    \centering
    \includegraphics[width=0.95\textwidth]{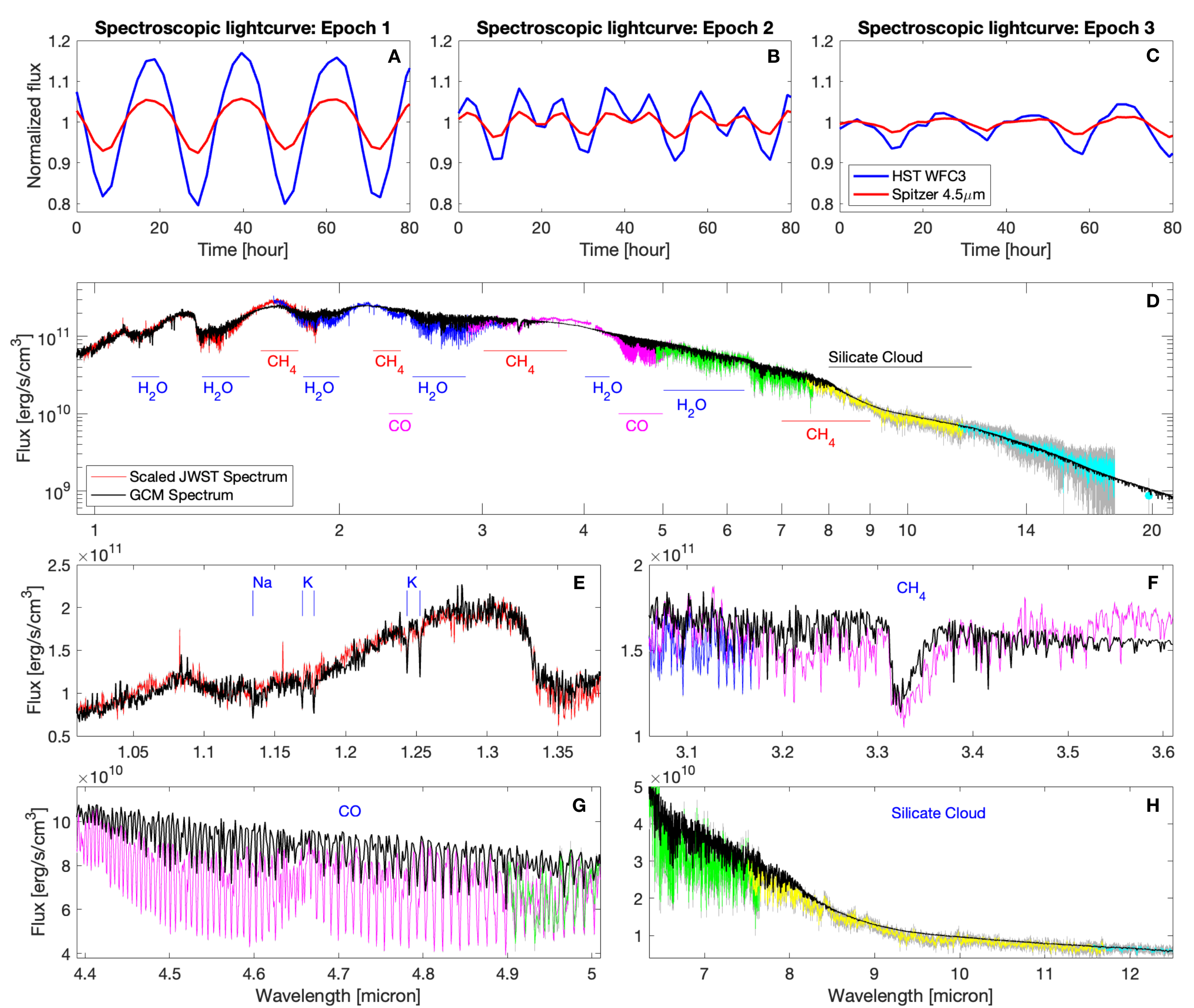}
    \caption{\textbf{Model spectroscopic light curves, spectrum, and comparisons to the JWST data.}
    {\footnotesize \textit{Panels A, B, and C:} synthetic light curves from the GCM in three epochs of outputs corresponding to three characteristic evolution stages of the equatorial dust storm. Blue lines are light curves running through the HST/WFC3 G141 filter and the red lines are those running through the Spitzer 4.5 $\mu$m filter. \textit{Panel D:}  model spectrum of VHS 1256B based on the GCM outputs at the time when the light curve reaches the minimum at epoch 1, with a spectral resolution of $R\sim 3000$ (black line), and the observed spectrum of VHS 1256B obtained by JWST (colored lines) \cite{miles2023}. In the radiative transfer post-processing, we have slightly altered the gas compositions of H$_2$O, CO and CH$_4$ taking into account the possible chemical disequilibrium caused by dynamical mixing \cite{miles2023}. The JWST spectrum is scaled for a direct comparison with the model spectrum, and different colors are those measured by different instruments, and the grey lines represent error bars. Important molecular gas and silicate cloud absorbers are highlighted. \textit{Panels E, F, G, and H:} zoom-in of the spectra in a few interesting wavelength regions: panel E contains H$_2$O and alkali metal absorption; panel F shows primarily the CH$_4$ absorption; panel G highlights the CO absorption; and panel H exhibits the silicate cloud absorption.}
    }
    \label{fig.summary}
\end{figure}

\subsection*{Spectroscopic light curves and spectrum}
Our simulations also quantitatively explain the wavelength dependence of the light-curve amplitude in the spectroscopic light curve observations. As shown in Fig. \ref{fig.summary}, our model predicts a maximum peak-to-peak variability of nearly 40\% for the light curve in epoch 1 at the HST/WFC3 band and nearly a 10\% amplitude at the Spitzer 4.5 $\mu$m band. These amplitudes are smaller at epochs 2 and 3. The variability amplitudes are generally higher at the shorter near-infrared wavelengths, where lower gas opacity allows the escaping flux to penetrate larger fractional heights between patchy clouds, resulting in larger variations of brightness temperatures (fig. \ref{fig.CF_amp}). At longer wavelengths, including the Spitzer 4.5 $\mu$m photometric band, the variability amplitudes are generally smaller. fig. \ref{fig.maps_and_lightcurve} displaces surface flux maps of the HST/WFC3 band and the Spitzer 4.5 $\mu$m band from the radiative transfer post-processing (see Supplementary Materials), showing the higher spatial contrast of the flux at near-infrared than at mid-IR wavelengths.  In addition to this general trend, the variability amplitude exhibits local maximums at a few wavelength regions (see fig. \ref{fig.CF_amp}). At near-infrared, the local maximums are located primarily at H$_2$O opacity windows. The amplitudes are also strong at the CH$_4$ absorption regions of about 3.3 and 7.8 $\mu$m mostly due to the chemical inhomogeneity associated with temperature variations (fig. \ref{fig.TPs}). While clouds are the drivers of circulation and inhomogeneity through radiative feedback, large temperature anomalies associated with the feedback and the corresponding chemical variations appear to affect the variability at longer wavelengths. Overall, our result explains the large observed amplitude (nearly 40\%) of VHS 1256B taken by HST/WFC3 derived from observations nearly 2 years apart.

Previous spectral observations using ground-based instruments revealed an exceptionally red near-infrared color of  VHS 1256B \cite{gauza2015,petrus2023}. The latest observed spectrum covering from 1 to 20 $\mu$m with exquisite precision and spectral resolution was taken by the JWST (\cite{miles2023}, also see Fig. \ref{fig.summary}). Clouds have long been proposed to redden the spectra starting from the L-type dwarfs \cite{suarez2023}, but the dynamical mechanisms of cloud reddening have not been directly tied to lightcurve variability.  Our model simultaneously explains the light curve variability and spectrum, as shown in the data-model spectral comparisons in Fig. \ref{fig.summary}. The dust storms significantly redden the spectrum and push the peak fluxes to molecular opacity windows near 1.6 and 2.2 $\mu$m, showing a good agreement with the observed full spectral energy distribution by JWST. A deep silicate absorption at $\sim10~\mu$m is well explained by MgSiO$_3$ clouds with a characteristic radius of 0.02$~\mu$m that are well mixed throughout the atmosphere. Certain depletion of CH$_4$ relative to chemical equilibrium explains the absorption features near 3.3 and 7.8 $\mu$m as suggested in \cite{miles2023}. {\add While our model provides a good comparison to the overall spectral redness and the 10 $\mu$m silicate absorption feature, mismatches persist, for example, in the L band centered around $3.5~\mu$m and in wavelengths between 4.4 and 5 $\mu$m where the modeled absorption lines of CO are shallower than the observed lines. These mismatches are likely due to imperfections in the cloud microphysical parameterization, which is a long-standing challenge for all atmospheric models \cite{helling2019}. In addition, multiple parameters, including varying gravity and non-solar chemical abundances, could contribute to the spectral shapes in the near-infrared bands \cite{gandhi2023bd,petrus2024}.  Insufficient cooling of the upper atmosphere by the lack of some cooling species may also contribute. Despite these imperfections, comparisons shown in Fig. \ref{fig.summary} are likely sufficient to validate that our complex 3D GCM (which is not dedicated to spectral fitting) captures the main processes shaping the spectrum of VHS 1256B.  A comparison between the spectra obtained by JWST, GCM, and the best-fit patchy-cloud radiative-convective-equilibrium model (which was designed for spectral fitting) in \cite{miles2023} is shown in fig. \ref{fig.bestfit2023} further demonstrates the satisfactory explanation of the GCM to the observed spectrum.}

Furthermore, we argue that to obtain sensible explanations simultaneously for both the spectrum and variability of VHS 1256B, the majority of silicate clouds should settle out relatively quickly as large particles with sizes of several microns, while a small fraction is transported to higher altitudes in the atmosphere as sub-micron particles (see Supplementary Materials for more discussion). If cloud particles are all composed of particles of about 1 $\mu$m, or larger particles of several microns, both the spectra and variability amplitudes would be qualitatively incompatible with the current observations. Examples of such failure models are shown in the Supplementary Materials. The reason is that, on the one hand, small particles are much easier to lift such that the reddening is too strong, and there is no cloud inhomogeneity to sustain the heating and cooling patterns for wave generation. On the other hand, large particles settle down efficiently and the opacity loading is small, such that feedback is weak and the reddening is insufficient. On the atmospheric dynamics side, the cloud cycles of only large particles will experience a relatively short intrinsic oscillation timescale \cite{tan2019} than the wave formation timescale, therefore also unfavorable to form and sustain the tropical waves. Both scenarios with too large or too small particles fail to explain the spectrum or the variability. {\add Lastly, we also highlight the importance of including Fe clouds in driving the global-scale waves and the large-amplitude light curves, as shown by a test case in fig. \ref{fig.summary_all} and discussed in the Supplementary Materials.}

\begin{figure}
    \centering
    \includegraphics[width=0.7\textwidth]{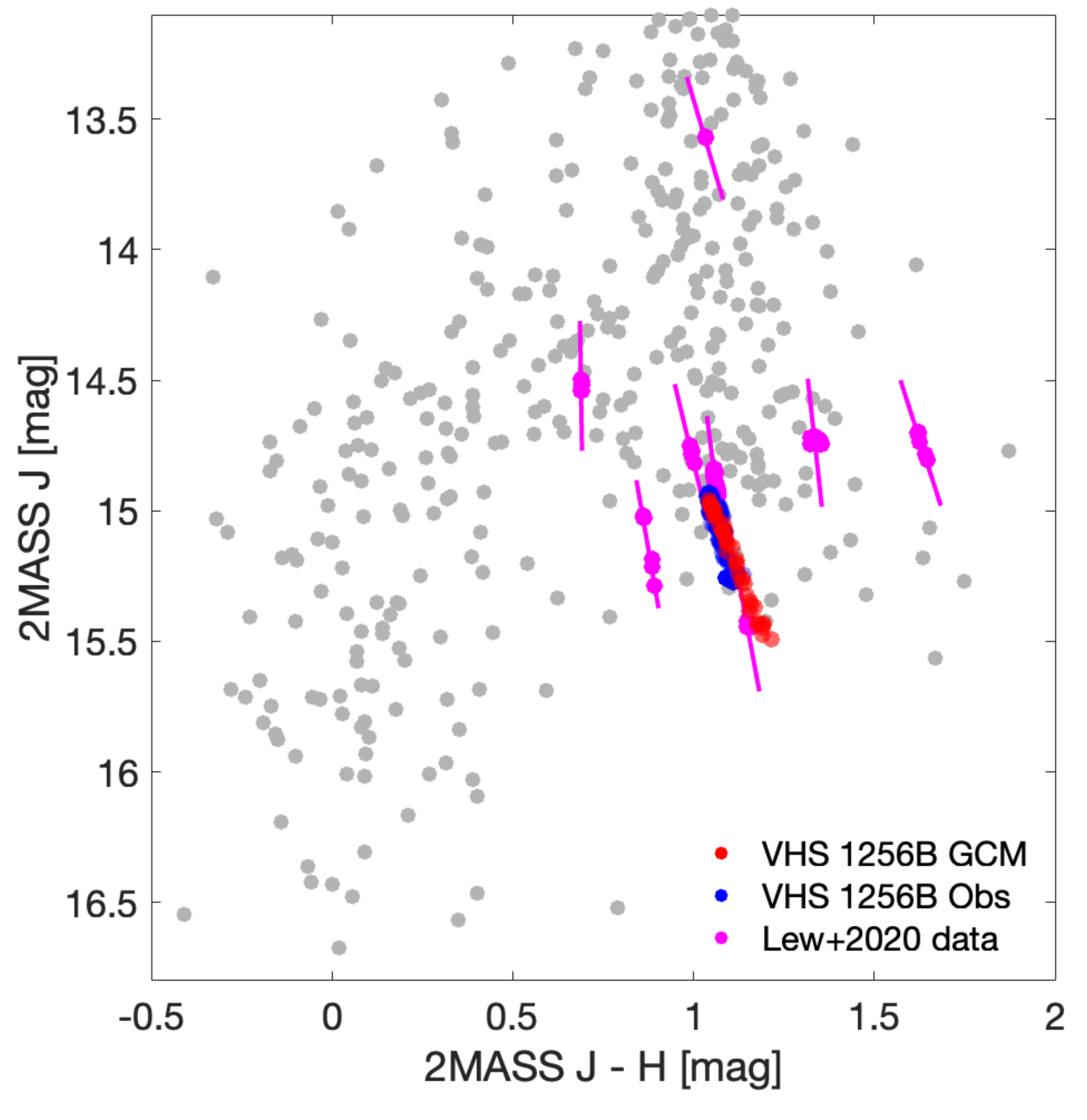}
    \caption{\textbf{Modeled variability in the near infrared $J$ vs $J-H$ CMD using 2MASS filters and comparisons to observed data.}
    The modeled variation trend represented by red points, based on time-resolved spectra at epoch 1 (Fig. \ref{fig.summary}), agrees well with that observed for VHS 1256B (blue points) using HST data of two epochs \cite{bowler2020,zhou2020,zhou2022}. Among other dusty brown dwarfs' variability observed with HST (magenta points), their variability trends in the CMD also qualitatively agree with our model outputs, indicating the same mechanism operates on a wider population of brown dwarfs and directly imaged giant planets. The magenta points include HST variability observations compiled in \cite{lew2020}, in which dots are observed time-resolved data and the lines represent linear fits to the data (see details in \cite{lew2020}). Note that in all colored points, the $J$ and $H$ band wavelength ranges are limited to 1.18 to 1.33 $\mu$m and 1.50 to 1.65  $\mu$m, respectively, as the original data observed by HST/WFC3 do not cover the full wavelength range of the 2MASS filters \cite{lew2020}. The grey dots are data obtained from the UltracoolSheet \cite{best_2024_10573247,dupuy2012,dupuy2013,liu2016bd,best2018,best2021,sanghi2023,schneider2023} whose wavelength ranges cover the full 2MASS filters. 
    }\label{fig.cmd}
\end{figure} 

\section{Discussion}
\subsection*{Implications for the {\add dusty substellar} population}
When placing VHS 1256B in the broader context of giant-planet and substellar atmospheres, our model offers a compelling explanation for the recently observed color variability trend in the near-infrared color-magnitude diagram (CMD) for dusty brown dwarfs and directly imaged giant planets (Fig. \ref{fig.cmd}). Observations from the HST/WFC3 variability survey \cite{buenzli2012,apai2013,buenzli2015,yang2015,lew2016,Manjavacas2018,biller2018,zhou2018,zhou2019,manjavacas2019, lew2020} revealed a strong correlation between the near-infrared J-H color variability and the J band magnitude as the object rotates. For L-type objects like VHS 1256B, they tend to become slightly bluer with increasing brightness during rotation. Our simulations explain this trend naturally: when the observed hemisphere of VHS 1256B is less cloudy, hotter flux escapes from a deeper atmosphere, leading to higher J-band magnitudes and shifting the spectrum blueward. Conversely, when more clouds are present, the spectrum becomes redder and flatter, and less flux is emitted at the cloud top, causing a dimmer near-infrared magnitude. This cloud-driven variability, particularly dominated by giant dust storms, explains both the observed color and magnitude changes in VHS 1256B (Fig. \ref{fig.cmd}). Given that similar trends have been observed in other cloudy exoplanets and brown dwarfs (Fig. \ref{fig.cmd}, \cite{lew2020}), our mechanism of cloud-induced variability could be universal among these {\add dusty} substellar populations. {\add Note that, while these atmospheres could share the same fundamental physical process---strong feedback between cloud cycles and atmospheric waves---the observational manifestation is expected to vary. The resulting weather pattern depends strongly on an object’s specific physical parameters, such as effective temperature, gravity, and rotation rate. For instance, the planet-sized storms that produce in-phase light-curve variations at different wavelengths are likely a consequence of slow rotation in our VHS 1256B simulation (fig. \ref{fig.fractional}). Different types of variability are present on faster-rotating brown dwarfs as shown in recent studies \cite{biller2024,chen2025}. }

Our 3D cloudy dynamics simulations of VHS 1256B, which feature planetary-scale giant dust storms rather than the traditionally assumed banded structures seen on solar system giants, provide a self-consistent framework for interpreting VHS 1256B’s observations. This framework not only reproduces VHS 1256B’s highly red spectra, but also explains four key aspects of its variability: large rotational modulation, irregular temporal evolution, wavelength dependence, and the correlation between color and magnitude variability in the near-infrared.  Our detailed study of VHS 1256B, one of the most well-characterized planetary-mass companions, provides valuable insights into the atmospheric physical, dynamic, and chemical processes of directly imaged giant planets and brown dwarfs. The cloudier low latitudes than the poles as a result of the cloud dynamics (fig. \ref{fig.clouds_zonal_mean}) naturally explain recent viewing-angle dependent observations of large samples \cite{vos2017,suarez2023}.   The abrupt transition from L to T-type dwarfs in the near-infrared CMD remains a long-standing puzzle in the spectral morphology evolution of substellar objects \cite{marley2015}. The cloud-hole scenario has compellingly explained the L/T transition \cite{ackerman2001,marley2010}, yet the dynamical processes to open up cloud holes remain poorly understood. This work shows that tropical waves can maintain large-scale cloud holes (Fig. \ref{fig.whitelight}) which is well benchmarked with observations of VHS 1256B, representing an important step towards resolving the mechanism triggering the L/T transition.

\section{Materials and Methods}

\subsection*{General Circulation Model}
The three-dimensional (3D) temperature, wind, cloud, and vapor structures of the substellar atmosphere are modeled using the SPARC/MITgcm general circulation model \cite{showman2009}. The model dynamical core solves the standard primitive equations of dynamical meteorology suitable for the observable atmospheres of giant planets and brown dwarfs \cite{vallis2017} using the cubed-sphere global grid based on the MITgcm \cite{adcroft2004}. The ideal-gas equation of state $p=\rho R T$, where $p$ is pressure, $\rho$ is gas density, and $R$ is the specific gas constant, of the hydrogen-dominated atmosphere is assumed. Radiative heating and cooling are the fundamental drivers of large-scale circulation. To obtain the radiative flux, the model solves the radiative transfer equations of planetary atmospheres based on a plane-parallel two-stream source function technique \cite{toon1989,marley1999,fortney2005}. The standard correlated-k method is utilized to obtain wavelength-dependent thermal fluxes following the set-up in \cite{kataria2013}. {\add We assume an instantaneous thermochemical equilibrium and one solar metallicity for the atmosphere. Dedicated spectral fitting work for VHS 1256B's JWST data showed no tight constraints for the atmospheric metallicity of VHS 1256B \cite{petrus2024}.  We chose a solar metallicity because it is a common and physically well-motivated baseline for young, L-type planetary-mass companions like VHS 1256b, for which there are no observational constraints suggesting a significantly non-solar composition. As a first-principle, complex 3D model, starting with a solar metallicity is the most reasonable and standard assumption in the field \cite{marley2015,showman2020}. Overall, the broader brown dwarf population is observed to have metallicities generally consistent with the solar value.} The opacity source in the radiative transfer includes the absorption by important molecules and atoms in giant planet and brown dwarf atmospheres, electron scattering, Rayleigh scattering of hydrogen and helium, as well as continuum opacity sources of collisional-induced absorption of H$_2$, H and He, bound-free and free-free opacity from major hydrogen and helium species (see a comprehensive list in \cite{marley2021}). The major difference in the radiative transfer between this work and previous work using the SPARC/MITgcm \cite{showman2009,kataria2013,parmentier2021,tan2024} is the lack of external stellar irradiation here. 

In the absence of pressure gradients induced externally by irradiation, the driving forces of large-scale circulation in self-luminous substellar atmospheres must come from internal heat. Recent work has proposed two robust mechanisms. The first mechanism concerns the interactions between the vigorous convection and the overlying stratosphere and the intersection between these injected energy and planetary rotation \cite{showman2013bd,zhang2014,showman2019,tan2022,hammond2023}. Despite the usefulness in accessing qualitative circulation properties, this mechanism relies on parameterized energy inputs, and the resulting light-curve variability is inevitably affected by the artificial parameter choices \cite{tan2022}. The second mechanism stems from the dynamic-radiative feedback driven by clouds. This was first discussed in the linear stability analysis for solar-system planets \cite{gierasch1973}, and Tan \& Showman \cite{tan2019} extended a similar concept in a one-dimensional model to capture more complexity of the cloud radiative feedback. Tan \& Showman \cite{tan2021bd1,tan2021bd2} explored this mechanism further in the nonlinear regime using self-consistent 3D atmospheric models, revealing the importance of planetary rotation in controlling the size of vortices, equatorial waves and the vertical extent of clouds. A similar methodology has been applied to more recent hot Jupiter and brown dwarf GCMs \cite{komacek2022,Lee2024bd}, and our current cloudy GCM framework has been used in preliminary modeling support for \cite{biller2024,Kleisioti2024,chen2024,chen2025}. 

In this model, we adopt the cloud radiative feedback mechanism to investigate the cloudy atmosphere of VHS 1256B. To do so, scalar tracer equations representing condensable vapor and clouds are simultaneously solved along with the dynamical core, as the setup in \cite{tan2021bd1,tan2021bd2}. We consider two cloud species most commonly considered in L-type dwarfs, ${\rm MgSiO_3}$ and Fe, in this work following \cite{ackerman2001}. For simplicity, we assume ${\rm MgSiO_3}$ consumes all Si elements and Fe droplets consume all Fe elements. Therefore, in the solar composition of \cite{lodders2003}, the mass mixing ratio of  ${\rm MgSiO_3}$ relative to H$_2$-He gas is about $2.3\times10^{-3}$ when all Si condenses out, and for Fe it is about $1.2\times10^{-3}$. Cloud particles are assumed to follow a log-normal size distribution $n(r) = \frac{\mathcal{N}}{\sqrt{2\pi} \sigma r}\exp\left(-\frac{[\ln(r/r_0)]^2}{2\sigma^2}\right)$ with prescribed mean radius $r_0$ and nondimensional width $\sigma$, where $r$ is the particle radius and $n(r)$ is the number density distribution. This size distribution was applied to substellar and planetary atmospheres since \cite{ackerman2001} to capture possible wide size distribution induced by condensation and coagulation, and has been widely used in the field of brown dwarfs and exoplanets (e.g., \cite{barman2011a,molliere2020}).  Clouds and vapor convert in between each other over a short timescale when the air parcels go across the condensation pressure-temperature (PT) profiles \cite{visscher2010}, and clouds also settle under gravity with speeds depending on the particle size, pressure, and temperature \cite{ackerman2001} --- the tracer transport, evaporation, condensation, and settling form a hydrological cycle in the atmosphere. The opacity loading of the location- and time-dependent clouds is included in the radiative transfer, giving rise to the radiative feedback that drives the atmospheric heating and cooling.

The deep layers in our GCM reach the convective zone that is typically at pressures larger than a few bars for L and T dwarfs \cite{burrows2001,marley2021}. In addition, regions where thick clouds form could also trigger local convection. The rapid convection and the mixing of entropy and tracers cannot be resolved in global models but can only be parameterized. We apply a simple convective adjustment scheme to instantaneously remove convective instability when it occurs and simultaneously mix tracers in the instability regions, following the prescription in \cite{collins2004}. Flow boundary conditions of the GCM are impermeable and stress-free. In the radiative transfer, we prescribe a uniform net upwelling internal heat flux characterized by the internal temperature $T_{\rm int}$. Depending on the planetary parameters, flow dynamics, and clouds, the internal heat flux will give rise to certain global mean thermal structures. Because the lower layers reach the convective zone whose entropy is supposed to be homogenized across the globe, we relax the temperature at the deepest pressure layer of the GCM towards a global mean state of that layer over a short timescale of $10^3$ s to represent the convective homogenization. This homogenization influences the rest of the column by convective adjustment and radiation. This treatment guarantees energy conservation while ensuring that each PT profile in the GCM is marched towards a single adiabatic state in the deep atmosphere. 

A linear drag is applied in the deep layers to represent momentum mixing between the GCM-resolved weather layer and the quiescent deep interior. The drag takes the form of $-k_v(p)\mathbf{v}$, where $p$ is pressure, $\mathbf{v}$ is horizontal velocity, and the drag coefficient $k_v(p) = \frac{1}{\tau_{\rm drag}} \max\left(0,\frac{p-p_{\rm drag,top}}{p_{\rm bot}-p_{\rm drag,top}}\right)$. The drag is strongest at the bottom pressure boundary $p_{\rm bot}$ and linearly decreases to zero at pressure $p_{\rm drag,top}$ with $\tau_{\rm drag}$ controls the overall dissipation strength. Any change in kinetic energy due to the drag is transferred {\add to} the local internal energy to ensure energy conservation. To minimize the direct impact of the drag on the observable layers, we set $p_{\rm drag,top}$ to 30 bars. $\tau_{\rm drag}$ is a free parameter and the nominal models in this work assume a strong drag with $\tau_{\rm drag}=10^4$ seconds. This is because the inner conducting regions of many observed L and T dwarfs are supposed to be close to the surface and effects on dragging the atmosphere should be strong. Nevertheless, we have carried out experiments with weaker bottom drag as discussed in the supplementary texts.   

Specific model parameters for VHS 1256 include an internal temperature of $T_{\rm int}=1150$ K and surface gravity of $g=190~{\rm ms^{-2}}$ based on the latest estimates in \cite{dupuy2023,miles2023}, the rotation period of 22 hours based on the Spitzer light curve in \cite{zhou2020}, the radius of the object of $7\times 10^7$ meters (nearly a Jovian radius).  The model pressure domain is from $10^{-4}$ bar to $50$ bars which are discretized to 53 layers in the vertical coordinate. The horizontal resolution is C64 of the cube sphere, equivalent to $256\times128$ in longitude and latitude, sufficient to resolve the dominant dynamical structures for a slow rotator like VHS 1256B. Most models were integrated for over 800 days where the mean atmospheric states have long been equilibrated, and those with longer $\tau_{\rm drag}$ (weaker bottom drag) were integrated for a longer time to reach statistical equilibrium. At statistical equilibrium, the global mean top-of-atmospheric thermal flux has a small ($\sim 1\%)$ fluctuation, meaning that the global mean state of the model is well-described by a single internal temperature.  Model outputs and post-processing were taken after the models reached statistical equilibrium. 

In our initial attempts, both  ${\rm MgSiO_3}$ and Fe clouds share a single size distribution that is held fixed throughout the domain and time, and we vary $r_0$ and $\sigma$ to explore the circulation, light-curve variability, and spectrum. Some of the attempts explained the near-infrared spectrum but failed to capture the spectral shape in the mid-IR range, and it was also difficult to simultaneously explain the large variability amplitude of VHS 1256B at near-infrared. 
Cloud microphysics models of substellar and exoplanet atmospheres have suggested very wide size distributions over a large range of parameter spaces \cite{helling2008b, gao2018,helling2019,gao2021,lee2023cloud,powell2024}. In reality, clouds at each size bin settle differently and are coupled differently to the gas dynamics, which is not captured by the {\add default GCM cloud parameterization that uses a single tracer to represent the full size distribution. This is because the settling flux is contributed mostly by large particles. When the size distribution is sufficiently wide, the quick settling of large particles will enforce the same quick settling for smaller particles (which in reality should not), inevitably because we model these clouds as ``a single mode" using one tracer.} 

It is too computationally expensive to perform the size-resolved cloud scheme \cite{gao2018,powell2024} in the GCM. To take a step forward in resolving the different couplings between clouds of varying sizes to the gas dynamics, we use different tracers to represent cloud particles of different characteristic sizes (but each tracer is still parameterized as the log-normal size distribution). We first add an additional tracer to represent a second size distribution mode of ${\rm MgSiO_3}$ clouds, which greatly improves the data-model comparison. In addition, the 10-$\mu$m silicate feature requires substantially small particles with size $\ll 1 \mu$m (e.g., \cite{cushing2006,hiranaka2016,luna2021}), for which we included a third mode of ${\rm MgSiO_3}$ with tiny particles to explain the 10-$\mu$m silicate absorption feature. This third tiny mode has little radiative feedback on the thermal structures and circulation because of its negligible opacity in near-infrared and mid-IR; therefore, it is essentially passive. {\add Having the first two cloud modes, with a large and a moderate mean size, also has to do with the mass budget of the silicate clouds. Given the assumption that all silicates will form condensates at appropriate conditions, the cloud mass per unit air mass must be constrained by the assumed atmospheric metallicity, which leads to quite massive clouds in terms of opacity. As discussed in the main text and in ``Additional results of the sensitivity tests" of Supplementary Text, the large mode serves to rain out a sufficient amount of clouds to avoid too much opacity in the atmosphere while a second, slow-settling mode is retained to trigger the global-scale waves as well as generating the appropriate redness of the spectrum.}

For the model presented in the main text, the first mode of ${\rm MgSiO_3}$ (also the same mode for Fe droplets) has $r_0=7~\mu$m and $\sigma=0.5$ with a typical mixing ratio $\sim 2.2\times10^{-3}$, the second mode of  ${\rm MgSiO_3}$ has $r_0=0.7~\mu$m and $\sigma=0.5$ with a characteristic mixing ratio $\sim 2\times10^{-4}$, and the third mode of  ${\rm MgSiO_3}$ has $r_0=0.02~\mu$m and $\sigma=0.3$ with a characteristic mixing ratio $\sim 2\times10^{-4}$.  Note Fe clouds are treated as only one size mode because they are typically formed deeper than the ${\rm MgSiO_3}$ clouds, having no direct impact on the spectrum. The supplementary texts present more discussion on the choice of cloud parameters and sensitivity tests. 

Lastly, we clarify the choice of ${\rm MgSiO_3}$ to represent silicate cloud but not other potential silicate species, {\add such as ${\rm Mg_2SiO_4}$ or ${\rm SiO_2}$}, as have been considered in other models \cite{visscher2010,helling2019,luna2021,Moran2024}. As the order-of-magnitude opacity is similar among different choices of silicate species, different choices of the silicate species may lead to subtle differences in the cloud-formation-dynamics-opacity feedback because they have slightly different condensation levels and material density. This could result in minor differences in cloud mixing and the resulting spectrum when other parameters (most importantly, cloud sizes) remain fixed. The redness of the spectra is mostly affected by the cloud continuum opacity, and the uncertainty of cloud choices may be folded into the uncertainty of cloud sizes, because the spectrum and dynamics care mostly about how much opacity, rather than what kinds of cloud contribute to the opacity, being mixed up in the atmosphere. {\add In the limit of efficient cloud vertical mixing that is preferred in the case of VHS 1256B, the cloud tops are in general much higher than the condensation level. This indicates that the emergent spectrum could mostly depend on the dynamics and may not directly link to light variations of silicate condensation levels. This points to possible degeneracies between particle sizes which mostly determine how well clouds are mixed up and the condensation levels which are determined by the condensation species. In this work with a primary focus on storm dynamics, we attempt to absorb this uncertainty by tuning the particle size distributions.} On the 10-micron absorption feature, the smallest ($\sim$ 0.02 micron) ${\rm MgSiO_3}$ or ${\rm Mg_2SiO_4}$ particles work similarly well, but not SiO$_2$, for which we have tested in our radiative transfer post-processing.  We also note that cloud particles in substellar and planetary atmospheres may be heterogeneous (i.e., individual particles are composed of a mixture of various substances) as predicted by {\add microphysics} models \cite{helling2019,gao2021} and have been proposed to explain recent spectral observations \cite{suarez-metchev2023,hoch2024}. But it {\add may} have a limited impact on the key results of this work because the dynamics and the first-order spectral reddening depend on the bulk cloud continuum opacity (rather than the subtle differences of the cloud opacity), as argued above.

\subsection*{Radiative Transfer Post-Processing: PICASO} 
Based on GCM 3D outputs of temperature and cloud structures, we calculate the time-dependent spectra using \texttt{PICASO}, an open-source radiative transfer code for computing the reflected, thermal, and transmission spectrum of planets and brown dwarfs \cite{batalha2019,Mukherjee2023}. The methodology of \texttt{PICASO} to calculate the radiative transfer for thermal emission is heritage from \cite{toon1989,marley1999}. Emission, reflection spectra, and phase curve calculations by \texttt{PICASO} based on the \texttt{SPARC/MITgcm} outputs of hot Jupiters have been carried out in a few studies \cite{Adams2022,Robbins2022,tan2024}.

The unique feature of the \texttt{PICASO-SPARC/MITgcm} coupling is that they share the same opacity source, the same technique of solving the radiative transfer equation (although using a different spectral resolution), and the same chemistry scheme. All these guarantee an energetically and chemically self-consistent treatment between the GCM outputs and post-processing, which is essential for using the spectrum to diagnose the effects of the 3D cloud cycle on the thermal spectrum. To compute spectroscopic light curves and phase-resolved spectra, we use the correlated-k method with 196 spectral windows over the whole wavelength range of 0.26 to 267 $\mu$m (equivalent to a resolution of $\sim 100$ in the near-infrared) in \texttt{PICASO} with pre-calculated k-coefficients and equilibrium chemistry. To compute a spectrum with higher resolution (as shown in Fig. \ref{fig.summary} and fig. \ref{fig.bestfit2023}) than that using the correlated-k method, we use the opacity re-sampling method in \texttt{PICASO} to first calculate a spectrum at a spectral resolution of $R\sim 50,000$, then bin it down to $R\sim 3,000$.  The computational efficiency of radiative transfer post-processing is much lower than the coarser radiative transfer calculations inside the GCM, and therefore we do not apply the full spacial grid resolution of the GCM to \texttt{PICASO} but downgrade the horizontal resolution to $61\times 61$ by longitude and latitude in the observable hemisphere (equivalent to about 3 degrees per cell) which is sufficient to capture important features.

{\add We have tested the opacity re-sampling post-processing based on additional original spectral resolutions of 100,000 and 200,000. We then bin the spectra down to a resolution of 3,000 and compare them to that obtained from the original spectral resolution of 50,000, focusing on 4.4 to 5 $\mu$m where CO absorption dominates. The overall depth of the absorption is not quite sensitive to the different original spectral resolutions. Minor discrepancies present but they are much smaller than either the overall absorption depth or the differences of data-model comparison. }

\bibliography{draft} 
\bibliographystyle{sciencemag}



%

%
%
%
%
%
%


\section*{Acknowledgments}
We acknowledge the computational support provided by the Siyuan-1 cluster, supported by the Center for High Performance Computing at Shanghai Jiao Tong University, and the Pleiades supercomputer at NASA.
\paragraph*{Funding:}
XT is supported by the National Natural Science Foundation of China (grant No. 42475131) and a Shanghai Jiao Tong University start-up grant. XZ is supported by the National Science Foundation Astronomy and Astrophysics Research Grants 2307463 and 2206317, the NASA Exoplanet Research grant 80NSSC22K0236, and the NASA Interdisciplinary Consortia for Astrobiology Research grant 80NSSC21K0597. YZ acknowledges HST data analysis grants associated with programs GO-16036. JMV acknowledges support from a Royal Society—Research Ireland University Research Fellowship (URF/R1/221932). 
\if false
\paragraph*{Author contributions:}
XT and XZ conceived this study. XT led the simulations, analysis, and manuscript writing. XZ contributed to the simulations and writing.  MM contributed to the interpretations of the simulations. YZ, BL, and BM provided the published observational data presented in this work. NB, MM provided help in simulations. All co-authors contributed comments and suggestions for the manuscript and active discussions of VHS 1256B's spectra and atmosphere.
\fi
\paragraph*{Author contributions:}
Conceptualization: XT, XZ, YZ, NW, ER, EM, AS, MM, KH, BB, BM, SM\\
Methodology: XT, XZ, YZ, ER, NB, KH \\
Software: XT, XZ, NB, MM, KH\\
Validation: XT, XZ, EM, SP, SH, GC\\
Formal Analysis: XT, XZ\\
Investigation: XT, XZ, YZ, NW, AS\\
Resources: XT, XZ, YZ, KH, BB\\
Data Curation: XT, YZ, EM, KH\\
Writing--original draft: XT, XZ, BS, GS\\
Writing--review \& editing: XT, XZ, JV, YZ, ER, NB, MM, KH, BB, SP, BS, GS, BP, SH, GC, SM\\
Visualization: XT, XZ\\
Supervision: XT, XZ, EM, AS\\
Project Administration: XT, XZ, AS\\
Funding Acquisition: XT, XZ, AS
\paragraph*{Competing interests:}
There are no competing interests to declare.
\paragraph*{Data and materials availability:}
All data needed to evaluate the conclusions in the paper are present in the paper and/or the Supplementary Materials.
The simulation data in this work are publicly available in Zenodo \url{https://doi.org/10.5281/zenodo.17213275}. \\
The \texttt{MITgcm} is publicly available at \url{https://mitgcm.readthedocs.io/en/latest/index.html} and \texttt{PICASO} is publicly available at \url{https://natashabatalha.github.io/picaso/index.html}.



\subsection*{Supplementary materials}
Supplementary Text\\
figs. S1 to S14


\newpage


\renewcommand{\thefigure}{S\arabic{figure}}
\renewcommand{\thetable}{S\arabic{table}}
\renewcommand{\theequation}{S\arabic{equation}}
\renewcommand{\thepage}{S\arabic{page}}
\setcounter{figure}{0}
\setcounter{table}{0}
\setcounter{equation}{0}
\setcounter{page}{1} 


\begin{center}
\section*{Supplementary Materials for\\ \scititle}

Xianyu Tan$^{\ast}$,
Xi Zhang,
Mark S. Marley,
Yifan Zhou,
Ben W. P. Lew,
Brittany E. Miles,
Natasha E. Batalha,
Beth A. Biller,
Gaël Chauvin,
        Sasha Hinkley,
        Kielan K. W. Hoch,
        Elena Manjavacas,
        Stanimir Metchev,
        Simon Petrus,
        Emily Rickman,
        Andrew Skemer,
        Genaro Suárez,
        Ben J. Sutlieff,
        Johanna M. Vos,
        Niall Whiteford
\end{center}

\subsubsection*{This PDF file includes:}
Supplementary Text\\
Figures S1 to S14\\


\newpage



\section*{Supplementary Text}

\subsection*{Additional results of the main model}\label{ch.s1}

\begin{figure}
    \centering
    \includegraphics[width=\linewidth]{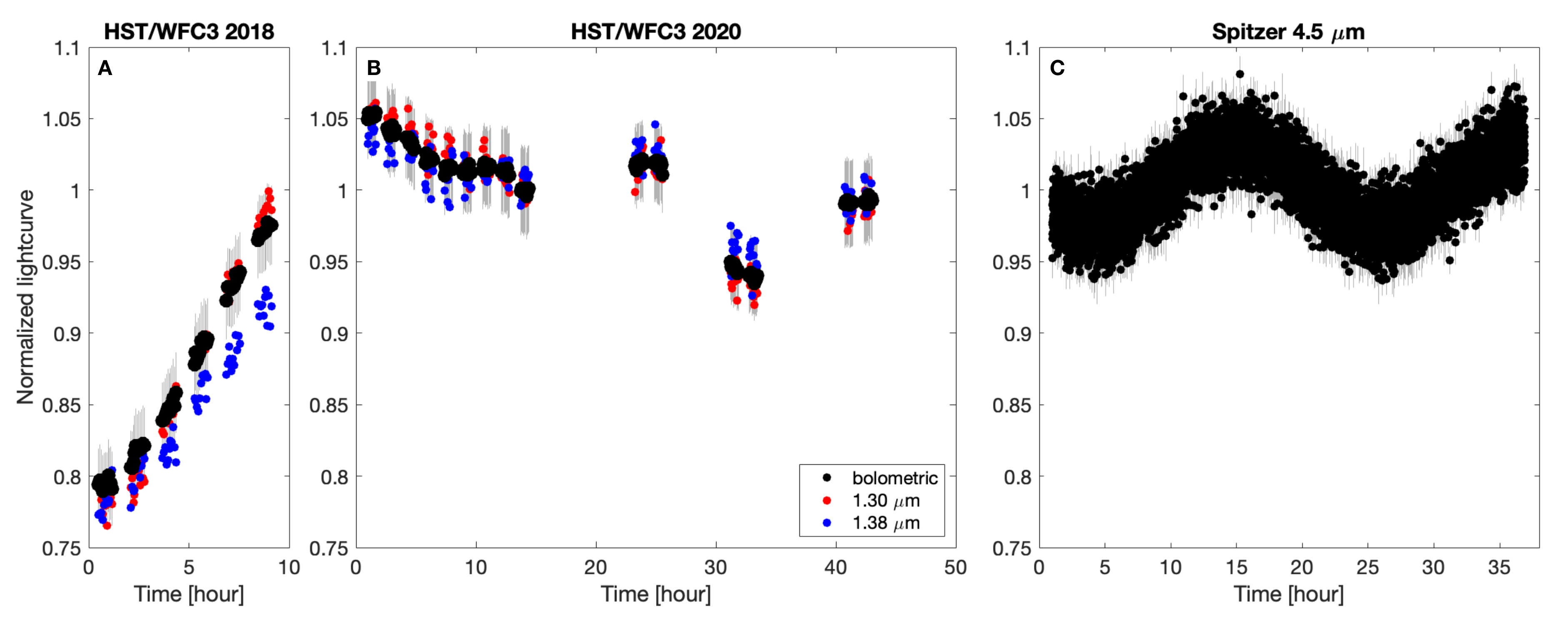}
    \caption{\textbf{Previously observed HST/WFC3 and Spitzer 4.5 $\mu$m light curves of VHS 1256B presented in \cite{bowler2020,zhou2020,zhou2022}.} The light curve of HST/WFC3 is normalized by the median value of the data obtained in both epochs. Black dots in the HST/WFC3 panels A and B are fluxes integrated through the bandwidth, red dots are fluxes at 1.30 $\mu$m (peak flux at the water opacity window), and blue dots are fluxes at 1.38 $\mu$m (low flux at the water absorption band). The light curve of Spitzer 4.5 $\mu$m is normalized by its median value. Grey lines are error bars of the band-integrated fluxes.
    }\label{fig.obs_lc}
\end{figure}

Fig. \ref{fig.obs_lc} summarizes light-curve observations of VHS 1256B, showing large variability amplitudes in near-infrared and mid-IR wavelengths obtained by HST and Spitzer. 

Here we describe additional results that support the main text. While it is difficult for our GCM to match the exact observed light-curve morphologies due to the turbulent and chaotic nature of the circulation, we aim to reproduce qualitative features of the light curves, including the amplitudes and inter-annual variations of the light curve features.

\begin{figure}
    \centering
    \includegraphics[width=0.7\linewidth]{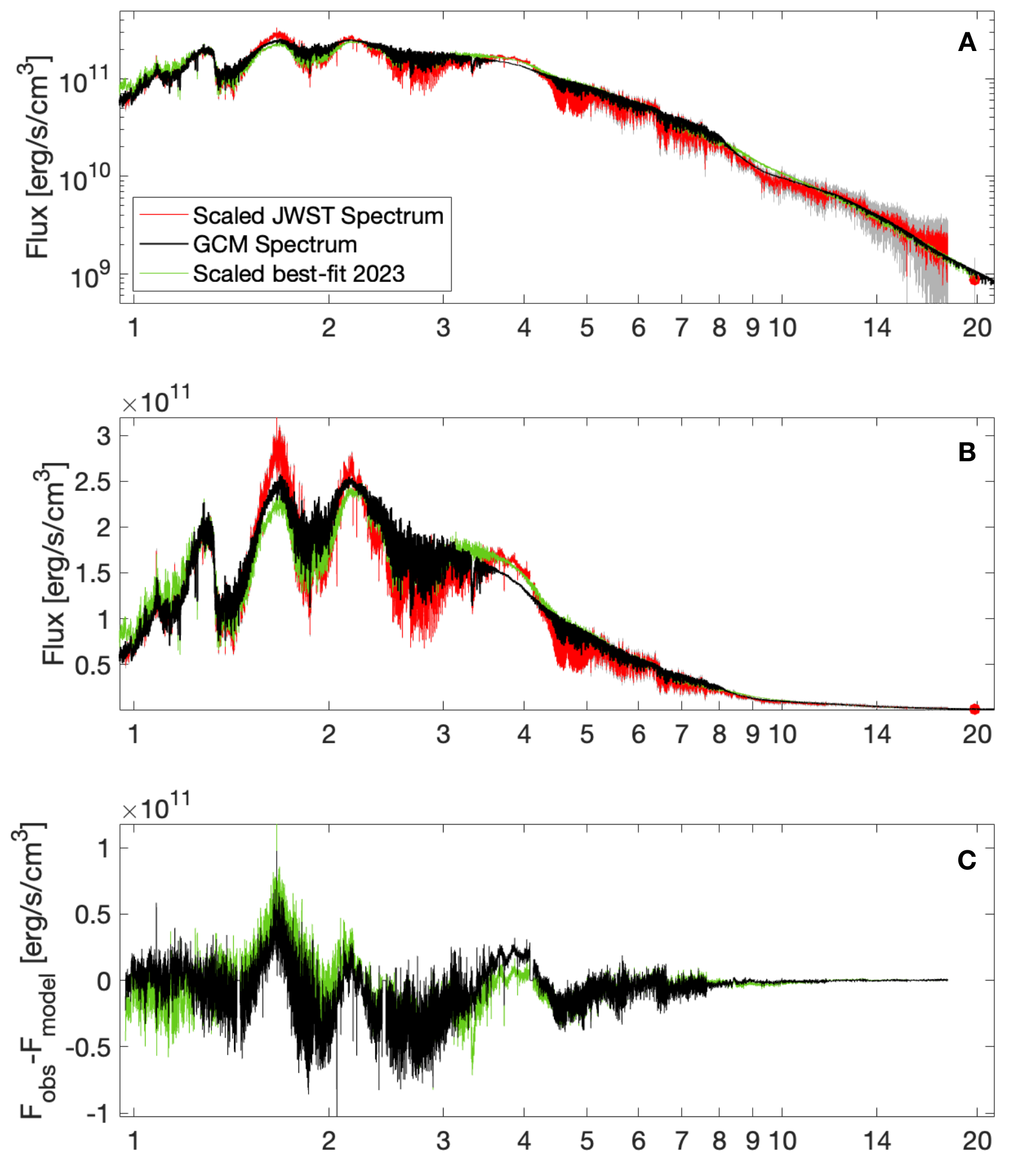}
    \caption{\textbf{Comparisons of spectra between the JWST observations (red), GCM (black), and the best fit based on a two-column radiative-convective model (green) presented in Miles et al. \cite{miles2023}.}  Panels A and B are shown in logarithmic and linear scale on the vertical axis, respectively.  Both modeled spectra have a spectral resolution of $R\sim 3,000$. Panel C shows the residues of two sets of data-model comparisons, in which modeled spectra are binned to the wavelength bins of the observed spectra.  The spectra are scaled such that they have the same integrated flux over the wavelength range covered by the JWST spectral observation (0.97 to 18 microns).
    }\label{fig.bestfit2023}
\end{figure}

Similar to Fig. \ref{fig.summary} in the main text, in fig. \ref{fig.bestfit2023}, we perform an additional comparison between spectra obtained from JWST, GCM and the best-fit 2-column radiative-convective-equilibrium model presented in the JWST data paper \cite{miles2023}. The redness is similar to the 1st order between the two types of models, although the GCM spectrum does a better job in the very blue part. The GCM spectrum {\add explains} the 10-$\mu$m silicate absorption feature better because the GCM includes MgSiO$_3$ particles with a small size ($\sim 0.02~\mu$m).

\begin{figure}
    \centering
    \includegraphics[width=0.8\linewidth]{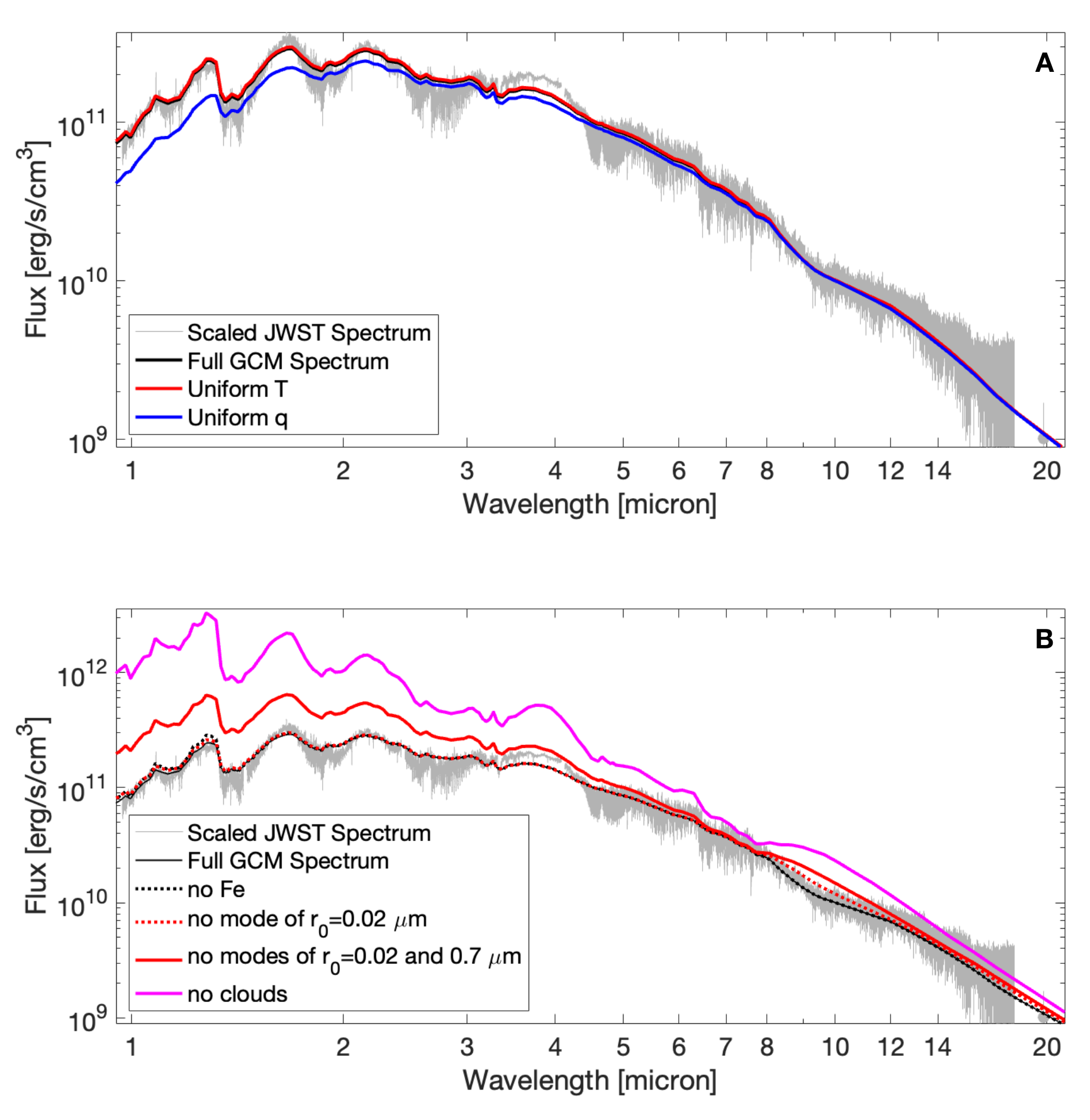}
    \caption{\add \textbf{Diagnoses of modeled spectra.} {\it Panel A:} scaled JWST spectrum, full GCM spectrum, a diagnostic spectrum (red) based on a globally uniform temperature structure, eliminating the effects of inhomogeneous temperature structures, and a diagnostic spectrum (blue) based on a globally uniform cloud structure, eliminating the effects of inhomogeneous cloud structures. {\it Panel B:} scaled JWST spectrum, full GCM spectrum, a diagnostic spectrum (black dotted line) excluding Fe cloud opacity, a diagnostic spectrum (red dotted line) excluding small silicate clouds of $r_0=0.02~\mu$m, a diagnostic spectrum (red line) excluding silicate clouds of $r_0=0.02$ and 0.7$~\mu$m, and a diagnostic spectrum (magenta line) excluding all clouds.}
    \label{fig.cloudtest}
\end{figure}

{\add Additional diagnostic tests of radiative transfer post-processing are performed and shown in fig. \ref{fig.cloudtest} to better clarify the roles of cloud and temperature inhomogeneity in affecting the modeled spectrum. In the GCM outputs, we artificially replace the 3D temperature field with a horizontally averaged temperature-pressure profile but keep the original 3D cloud field. The resulting post-processed spectrum (represented by the red line in panel A) is almost identical to the original one. On the other hand, we replace the 3D cloud field with a horizontally averaged cloud-pressure profile but keep the original 3D temperature field, and this leads to a spectrum with significantly lower flux and redder near-IR colors (blue line). This test shows that the horizontally inhomogeneous cloud structure plays a much bigger role in the spectrum than the inhomogeneous temperature structure. This makes sense because the brightness temperature variations are more sensitive to the large variation of cloud-top pressures than the horizontal temperature variations in conditions relevant to VHS 1256B.  }

{\add We also test the influences of various cloud components in the spectrum as shown in  panel B of fig. \ref{fig.cloudtest}. In the GCM outputs, we artificially remove one or more cloud components to obtain the post-processed spectra shown in  panel B. When removing Fe clouds only, the spectrum has little difference to the original spectrum, indicating that Fe clouds are not directly visible from the spectrum because they form deeper than the silicate clouds. When removing the silicate mode with $r_0=0.02~\mu$m, the spectrum shows only differences in wavelengths larger than about 8 $\mu$m, because they have negligible opacity at shorter wavelengths. In the tests of removing silicate modes with $r_0=0.02$ and $0.7~\mu$m and removing all clouds, spectra have much higher fluxes than the original one, suggesting that silicate clouds with moderate and large particle sizes contribute to the opacity that reddens the out-going spectrum.  }

\begin{figure}
    \centering
    \includegraphics[width=\linewidth]{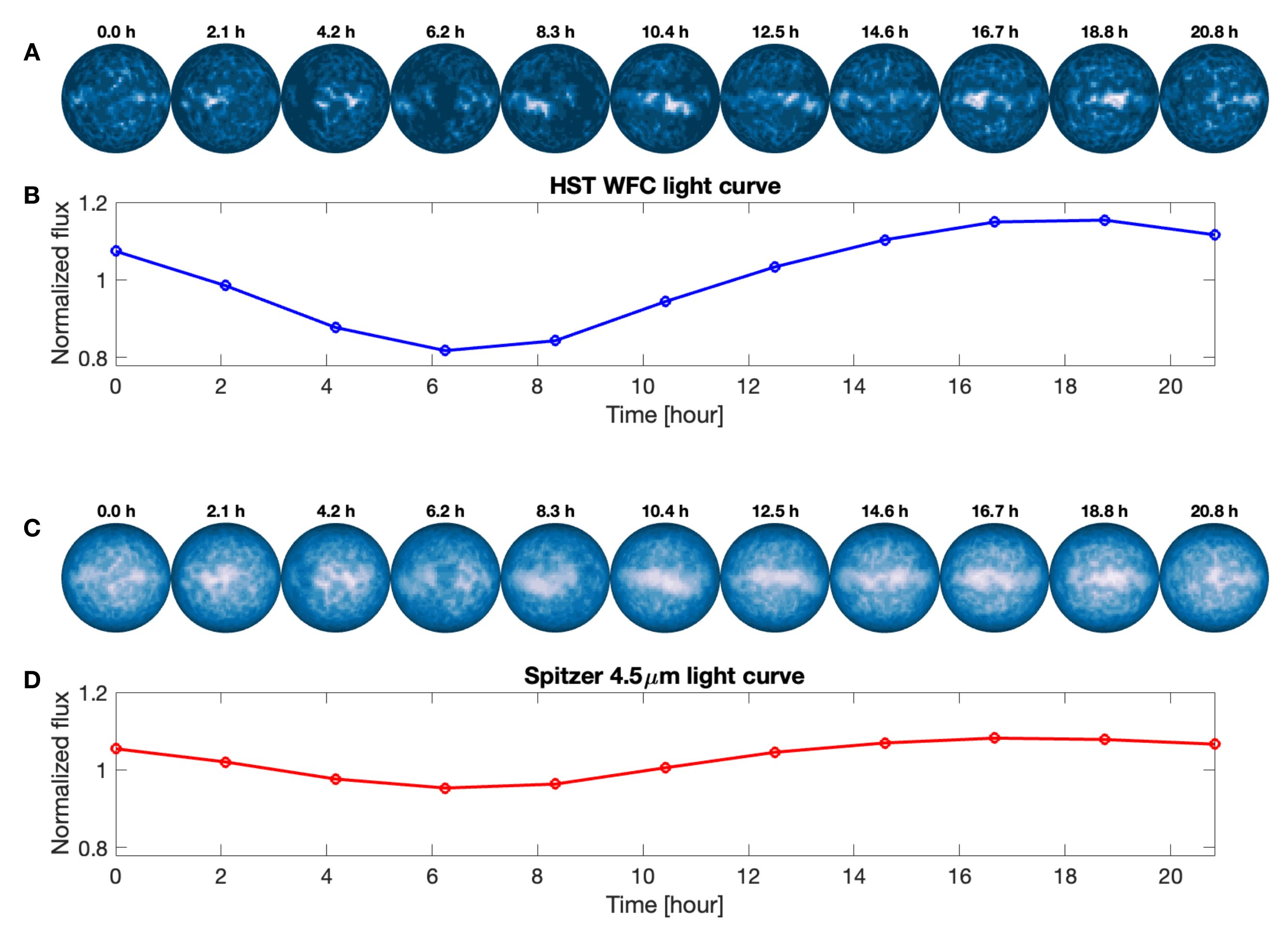}
    \caption{\textbf{Spatial flux maps outputted from \texttt{PICASO} post-processing based on GCM outputs, showing that the inhomogeneity comes mostly from the equatorial storm.} The spatial contrast of the maps is larger in the HST/WFC3 maps than in the Spitzer 4.5 $\mu$m maps, corresponding to the larger light-curve variability amplitudes in the HST/WFC3 band.  
    }\label{fig.maps_and_lightcurve}
\end{figure}

\begin{figure}
    \centering
    \includegraphics[width=0.9\linewidth]{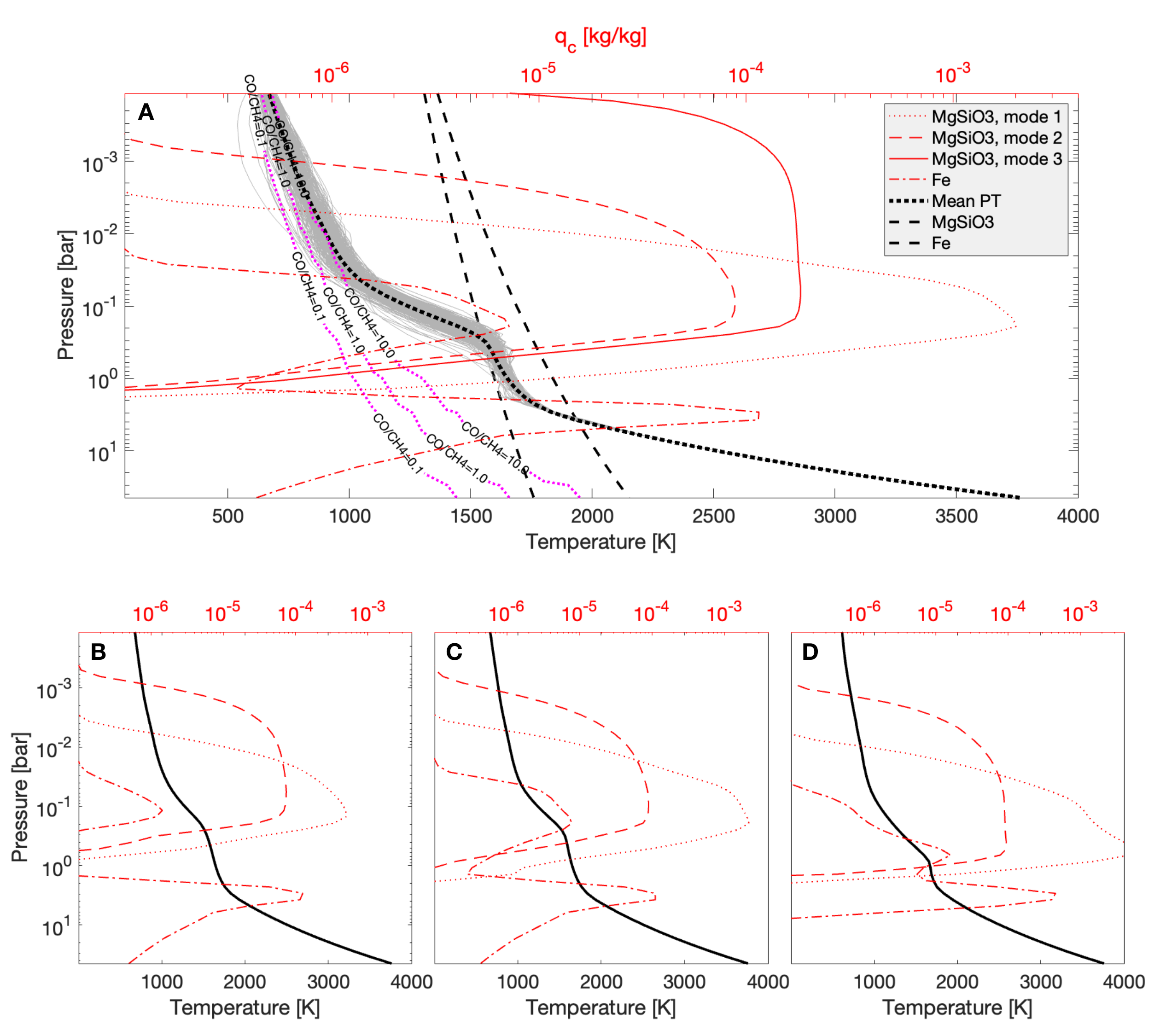}
    \caption{\textbf{Temperature and cloud profiles of the GCM.} {\add {\it Panel A}  contains randomly selected temperature-pressure profiles (grey lines) at an instantaneous output of the model, along with time-global-mean TP (thick dotted line), different cloud components (red lines), CO/CH$_4$ ratio (magenta lines) profiles, and condensation curve of MgSiO$_3$ and Fe (dashed lines). Mode 1 of MgSiO$_3$ refers to the mode with $r_0=7~\mu$m and $\sigma=0.5$, mode 2 refers to the mode with $r_0=0.7~\mu$m and $\sigma=0.5$, and mode 3 refers to the mode with $r_0=0.02~\mu$m and $\sigma=0.3$.  {\it Panels B, C, and D:} mean temperature and cloud profiles for the relatively cloud-free areas (left), medium cloud thickness area (middle), and cloud-thick area (right).}
    }\label{fig.TPs}
\end{figure}

Fig. \ref{fig.CF_amp} shows the global-mean contribution function $|dF_{\lambda}/dT|$ as a function of wavelength and pressure based on one instantaneous output of the GCM and the post-processing in  panel A, and the maximum deviation as a function of wavelength of the normalized light curve of epoch 1 (see Fig. \ref{fig.summary}) as a function of wavelength in panel B. The deviation generally decreases with increasing wavelength but shows local maxima in near-infrared regions of J, H and K bands where ${\rm H_2O}$ gas absorption is weak and in mid-IR regions of about 3.3 and 7.8 $\mu$m where ${\rm CH_4}$ absorption is strong. The overall decreasing trend of the amplitude with wavelength can be recognized from the global mean contribution function. As wavelength increases, fluxes escape from lower pressures where brightness temperature contrasts between cloud patches become smaller. The spatial thermal maps at different wavelengths produced by \texttt{PICASO} are shown in fig. \ref{fig.maps_and_lightcurve}, which helps to visualize the characteristic spatial contrast at near-infrared and mid-infrared wavelengths.   The ${\rm H_2O}$ absorption windows contrast more due to the weaker gas absorption and exhibit larger variability amplitudes. The larger amplitudes in the strong ${\rm CH_4}$ absorption bands are interesting, and they are associated with the horizontal chemical variations in which the gas mixing ratios of CO and CH$_4$ vary substantially. This is perhaps a coincidence that the mean PT of the photosphere lies near where CO-CH$_4$ transitions are significant and temperature variations driven by cloud radiative dynamics induce strong horizontal CO-CH$_4$ variations (see fig. \ref{fig.TPs}). Note that this effect is present in the chemical equilibrium assumption made in our GCM. Potential chemical disequilibrium induced by rapid gas mixing in the atmosphere may modify the quantitative features near these ${\rm CH_4}$ absorption bands \cite{miles2023,lee2023bd,Lee2024bd}. It is also interesting to note that there is no local minimum or maximum in the variability amplitude in the strong silicate absorption band of $\sim 10~\mu$m, in contrast to previous thoughts (e.g., \cite{luna2021}). This is because the smallest MgSiO$_3$ particles are well mixed in the photosphere due to the low settling speed and their patchiness is low. The variability at the $\sim 10~\mu$m mostly traces thermal variations that show the same characteristic behaviors as those in the nearby gas absorption wavelengths. 

\begin{figure}
    \centering
    \includegraphics[width=\linewidth]{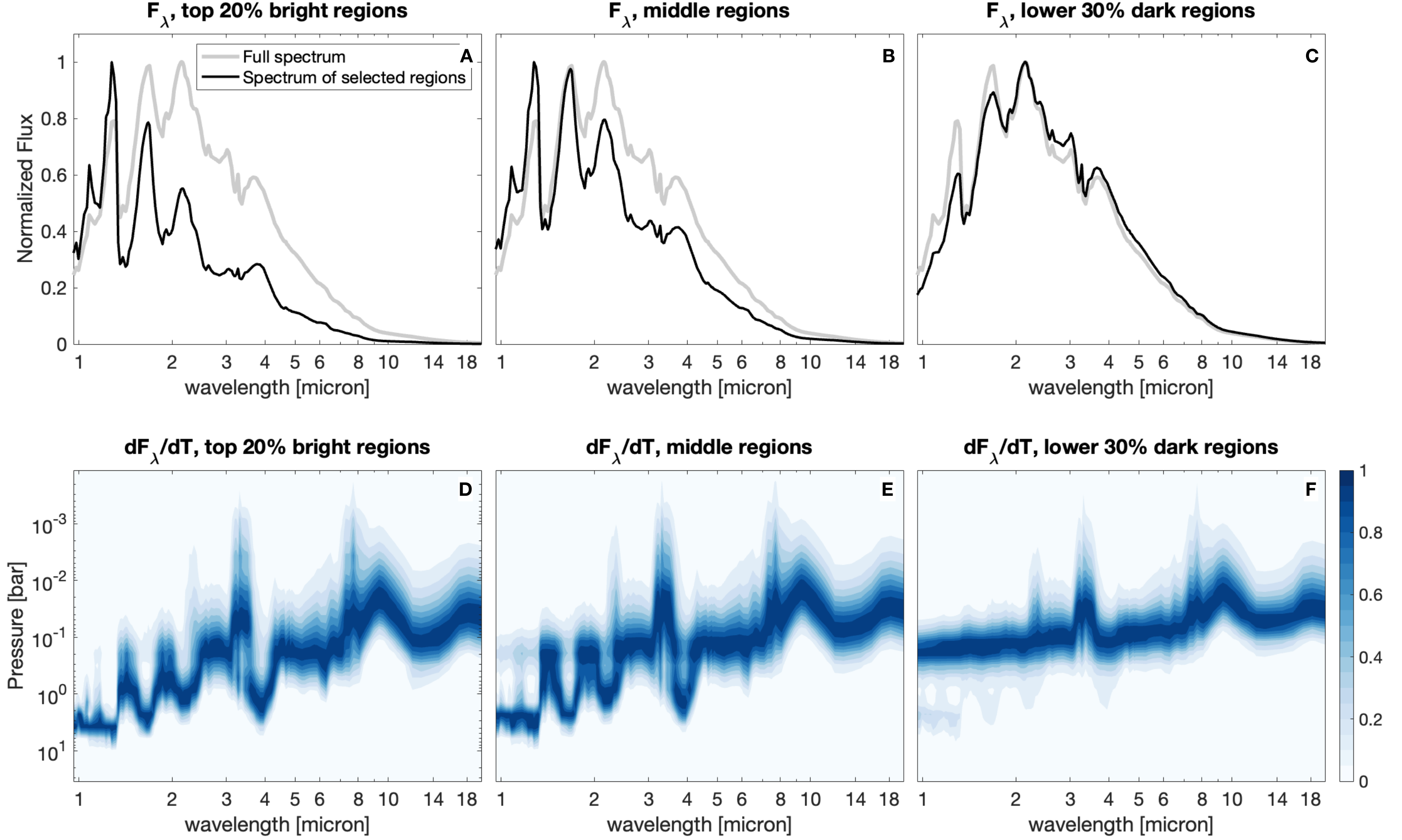}
    \caption{\add \textbf{Spectra and contribution functions from different regions of the GCM based on PICASO outputs.} {\it Panels A, B and C:} normalized spectral flux $F_{\lambda}$ summed over the top 20\% bright regions evaluated by flux at 1.2 micron (A), medium regions (B), and the lower 30\% dark regions (C). The grey, thick line represents the full spectrum summing over all regions, plotted for comparison. {\it Panels D, E, and F}: contribution functions $dF_{\lambda}/dT$ as a function of wavelength and pressure corresponding to the spectra shown in panels A, B, and C. 
    }\label{fig.CF_more}
\end{figure}

{\add Horizontal inhomogeneity of the atmosphere significantly influences the integral spectrum as shown in fig. \ref{fig.cloudtest}, and we further analyze local spectral properties that emerge from different regions of the atmosphere. In the PICASO flux map at 1.2 $\mu$m (which probes the deepest layers) similar to that shown in fig. \ref{fig.maps_and_lightcurve}, {\addmore ``bright regions" are defined as areas having the top 20\% local fluxes at 1.2 $\mu$m, i.e., fluxes larger than $F_{\rm min}+(1-20\%)\times(F_{\rm max}-F_{\rm min})$, where $F_{\rm max}$ and $F_{\rm min}$ are the maximum and minimum fluxes in the 1.2 $\mu$m flux map. These bright regions represent thin cloud regions. Conversely, ``dark regions" are referred to as areas having the lower 30\% local fluxes at 1.2 $\mu$m, representing regions of thick clouds. ``Middle regions" means the rest areas.} The normalized integral spectra of the three characteristic regions are shown in panels A, B, and C of fig. \ref{fig.CF_more} together with a comparison to the full GCM spectrum. Not surprisingly, the integral spectrum of bright regions is much bluer than either that of the other regions or the full spectrum. There is a nearly continuous transition of near-IR redness from the bright to dark regions. The corresponding contribution functions are shown in panels D, E, and F of fig. \ref{fig.CF_more}, demonstrating that the near-IR redness is mainly affected by variations of the local cloud top pressures. Recent atmospheric retrieval work using two columns with different cloud depths started to capture these strong horizontal variations of spectral properties \cite{vos2023,zhangzz2025}.    }

\begin{figure}
    \centering
    \includegraphics[width=\linewidth]{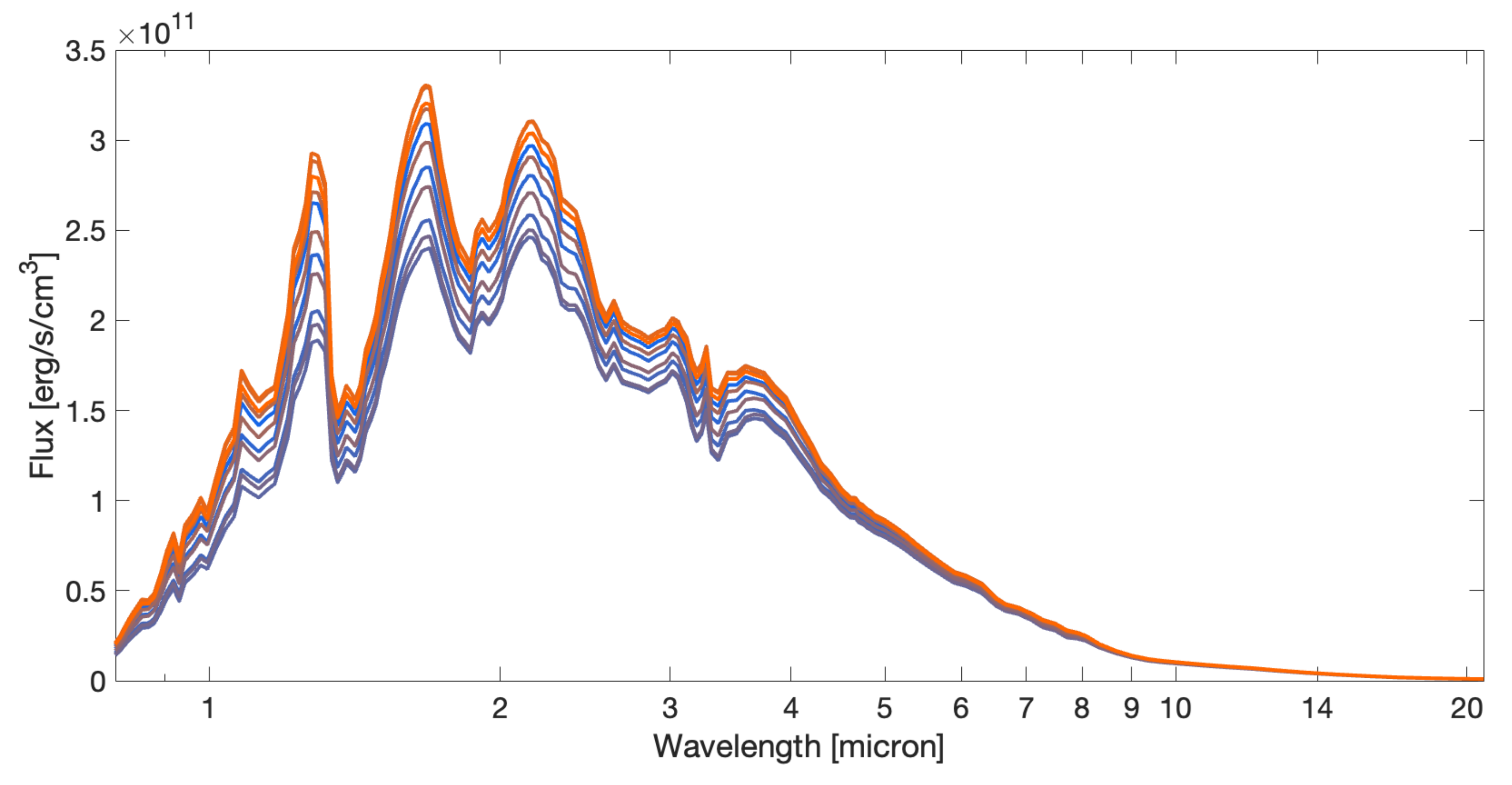}
    \caption{\textbf{Phase-dependent spectrum from the first rotation period of epoch 1.}
    }\label{fig.spectrum_time}
\end{figure}

\begin{figure}
    \centering
    \includegraphics[width=\linewidth]{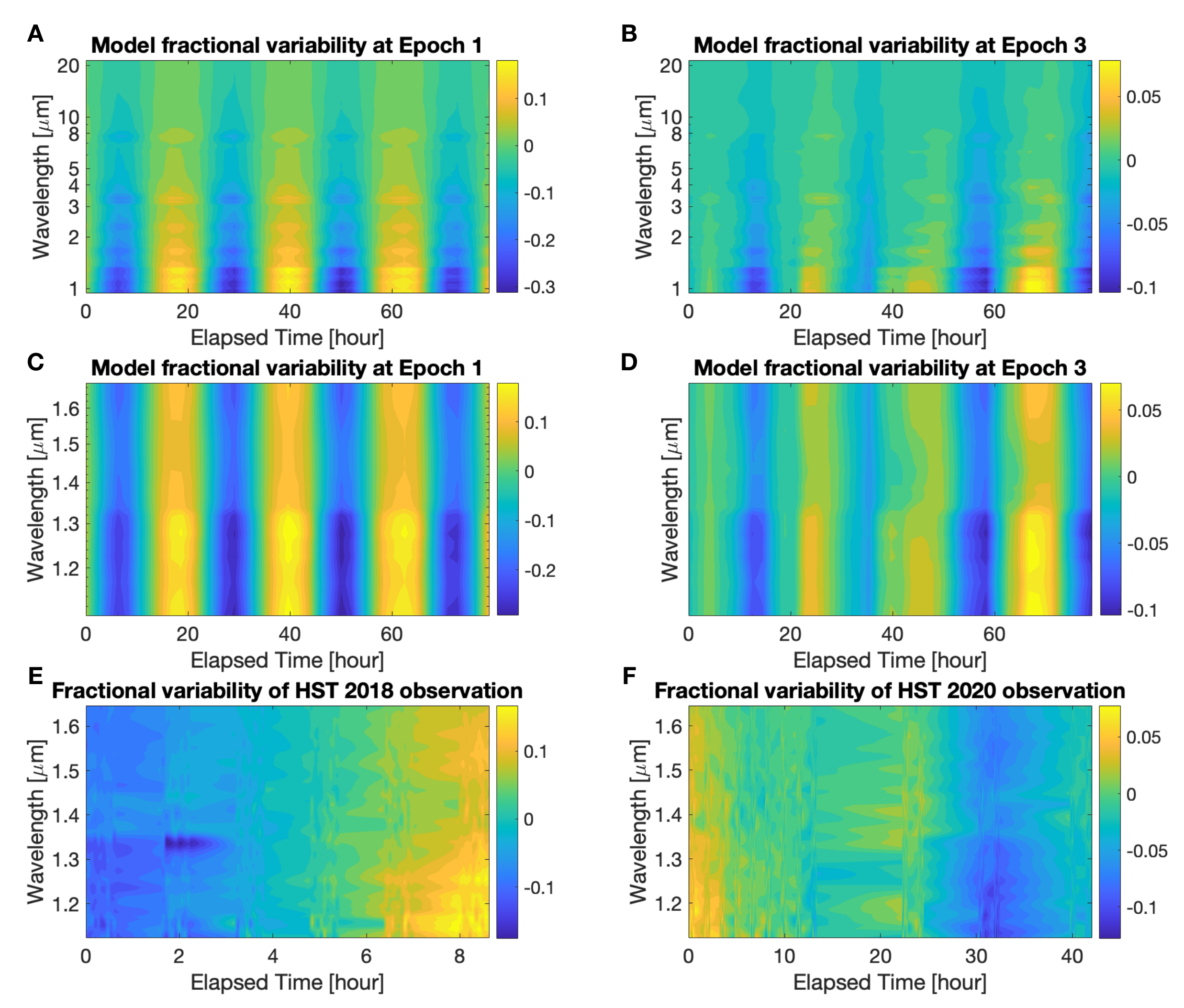}
    \caption{\textbf{Maximum deviation of the spectroscopic light curves as a function of wavelength and time for Epochs 1 and 3 shown in Fig. \ref{fig.summary}.} {\add {\it Panels A and B:} maximum deviation over 1 to 20 microns. It exhibits nearly in-phase variations across a wide range of wavelengths in both epochs. {\it Panels C and D:} a zoom-in of the upper panel to the HST/WFC3 wavelength range of 1.0 to 1.7 micron. {\it Panels E and F:} HST observation.  }
    }\label{fig.fractional}
\end{figure}

\begin{figure}
    \centering
    \includegraphics[width=\linewidth]{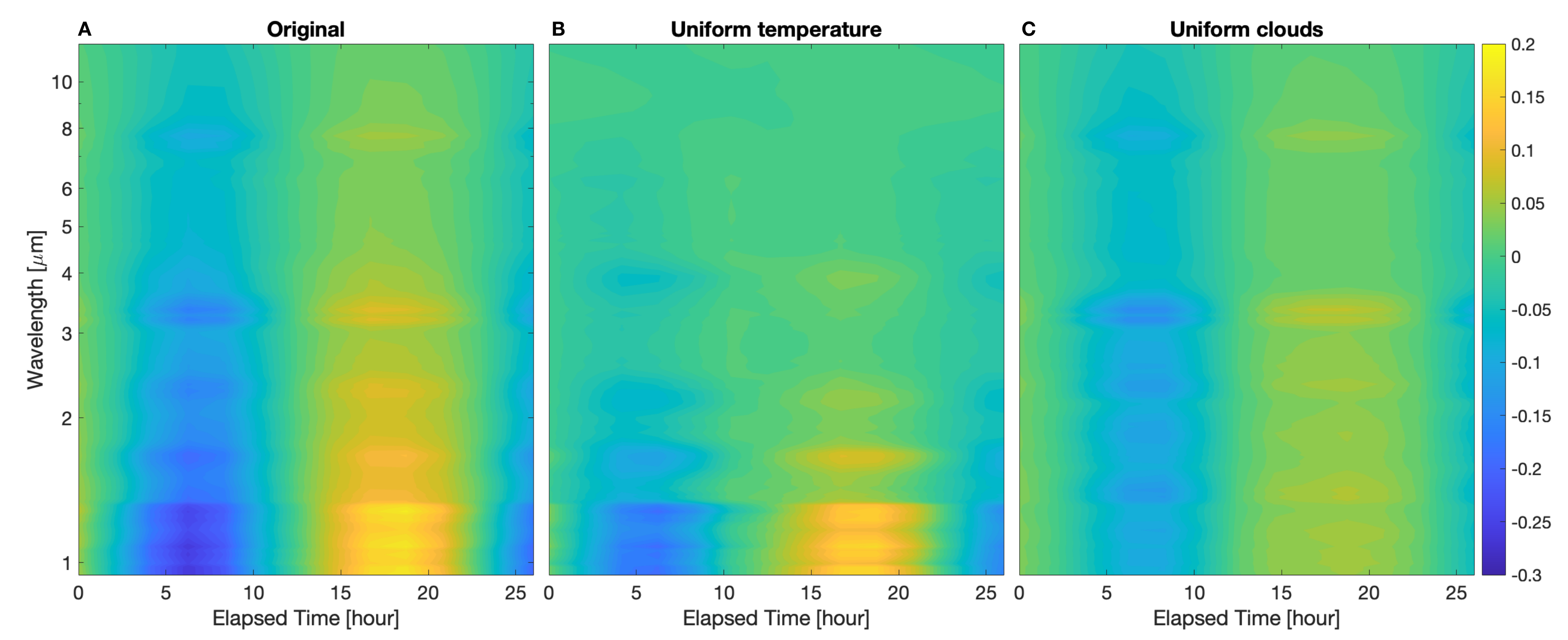}
    \caption{{\add \textbf{Diagnostic maximum deviation of the spectroscopic light curves as a function of wavelength and time for Epoch 1 shown in Fig. \ref{fig.summary} to clarify the role of inhomogeneous clouds and temperature in shaping the spectroscopic light curves.}  {\it Panel A:} normalized maximum deviation over 1 to 12 microns from the original light curves. {\it Panel B:} normalized maximum deviation from light curves based on a globally uniform temperature structure, eliminating the effects of inhomogeneous temperature structures in the light curves. {\it Panel C:} normalized maximum deviation from light curves based on a globally uniform cloud structure, eliminating the effects of patchy clouds in the light curves.   }
    }\label{fig.fractional_diag}
\end{figure}

Fig. \ref{fig.spectrum_time} contains the simulated spectra at different phases of a single rotation in the Epoch-1 light curve (Fig.\ref{fig.summary}). {\add fig. \ref{fig.fractional} shows the maximum deviation as a function of wavelength and time for the modeled light curves in epochs 1 and 3, from 1 to 20 $\mu$m in panels A and B and a zoom-in in the HST/WFC3 wavelength range in panels C and D, along with the maximum devation of those observed by HST in panels E and F.} The main feature is the alignment of the variability over all wavelengths, i.e., little phase offsets between different wavelengths. This is a consequence of the major circulation pattern being consistent over a large pressure range. This differs from previously reported light-curve observations of other objects (e.g., \cite{yang2016,biller2018,plummer2024}) {\add but is consistent with the HST spectroscopic light-curve observations of VHS 1256b \cite{zhou2020,zhou2022} as shown in panels E and F of fig. \ref{fig.fractional}}. {\add Both horizontal cloud and temperature variations should contribute to the spectroscopic light-curve variations, and now we further diagnose their relative roles. Similar to the exercise done for spectral diagnosis in fig. \ref{fig.cloudtest}, we artificially remove horizontal temperature or cloud variations of the GCM outputs, then calculate the spectroscopic light curves. The maximum deviations as a function of wavelength and time are compared to the original one shown in fig. \ref{fig.fractional_diag}. Both cloud and temperature variations produce in-phase variation across different wavelengths. However, in this case, cloud variations dominate the light-curve variations at a wide range of wavelengths, especially at near-IR. Temperature variations are important at wavelengths where molecular opacity is prominent, e.g., {\addmore the} CH$_4$ bands at about 3.3 $\mu$m. }

{\add Clouds shape the thermal structures substantially via the cloud greenhouse effect. Panel A in fig. \ref{fig.TPs} shows randomly selected PT structures, the global-mean mixing ratio of different cloud components, lines corresponding to CO/CH$_4=0.1,1$ and 10, and condensation curves of MgSiO$_3$ and Fe. The temperature lapse rates $d\ln T/d\ln p$ below the silicate cloud bases at near 1 bar are significantly lowered because of the cloud back warming \cite{ackerman2001, burrows2006,tan2019}. Panels B, C and D exhibit temperature and cloud profiles averaged over relatively cloud-free areas (left), medium cloud thickness areas (middle), and cloud-thick areas. Temperature structures of all areas are similarly affected by cloud opacity with moderate differences. Cloud vertical structures are also similar among these areas, with the main differences in the overall cloud column mixing ratios.   }

\begin{figure}
    \centering
    \includegraphics[width=\linewidth]{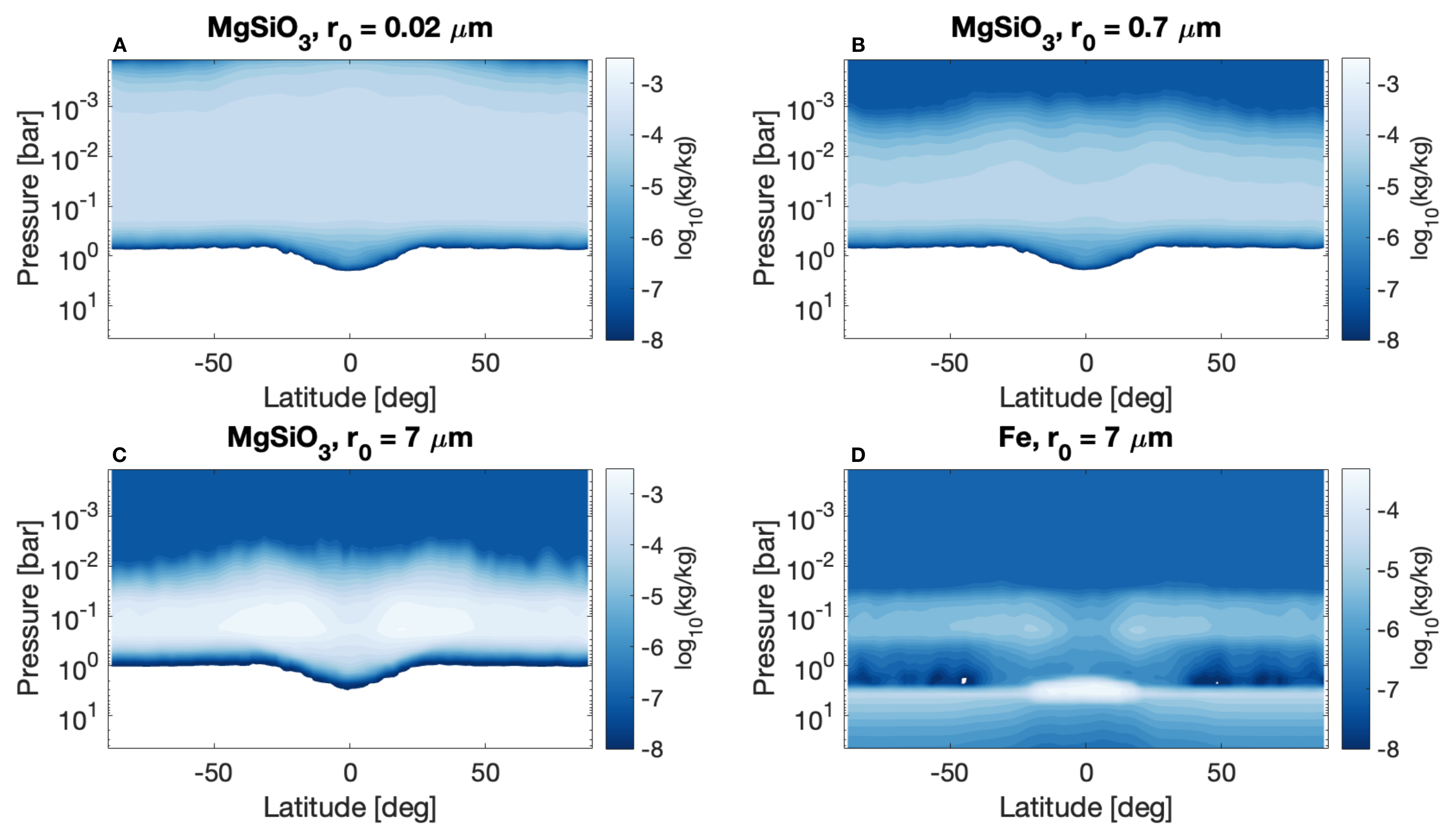}
    \caption{\textbf{Zonal-mean mass mixing ratios of different cloud components in the GCM.}
    }\label{fig.clouds_zonal_mean}
\end{figure}

In the model presented in the main text, despite the strong bottom drag at pressures larger than 10 bars, a prominent and broad westward jet developed at the equator and two narrower eastward jets at mid-latitudes, as shown in panel A of fig. \ref{fig.maps}. The westward direction of the equatorial jet is consistent with those previously investigated in brown dwarf GCMs \cite{showman2019,tan2021bd2,tan2022}. However, horizontal fields of zonal velocity, temperature and clouds {\add (for example, at surfaces of 0.01 bar, 0.12 bar, and 1.09 bar which bracket the majority of the main emission pressure layers, as shown in panels from C to K of fig. \ref{fig.maps})} do not exhibit a direct signature of the jets but the equatorial wave structure and the associated cloud distributions dominate their first-order inhomogeneity. The time and zonal-mean distributions of the cloud components of the model, as shown in fig. \ref{fig.clouds_zonal_mean}, display interesting global variations. Clouds are generally thicker and transported higher toward low latitudes, and this is primarily regulated by the planetary rotation \cite{tan2021bd1,tan2021bd2}. Observations of inclination-dependent near-infrared colors and silicate absorption features of brown dwarfs also support this feature \cite{vos2017,suarez2023}.  The subtropical regions (about $\pm 25$ degrees at latitudes) sustain the strongest vertical transport rather than at the equator. The subtropical regions correspond to where the strong off-equatorial Rossby gyres persist inside which the warm and cyclonic high-pressure centers support long-lasting cloud formation.  At the equator, the returning and subsiding flow from the Matsuno-Gill wave tends to sustain a zonally extended and relatively clear-sky stripe (see a long and relatively bright strip at the equator in panel A of Fig. \ref{fig.whitelight}, as well as the cloud map in fig. \ref{fig.maps}). As such, the zonal-mean vertical cloud transport tends to be weaker at the equator than in the subtropical regions. The zonal-mean condensation level can be straightforwardly visualized as the base of these cloud layers in fig. \ref{fig.clouds_zonal_mean} which is controlled by evaporation. Even though clouds appear to be  thick in the mean (either globally or in the zonal average), because clouds are horizontally inhomogeneous, the redness of the spectrum is contributed by both the relatively cloud-free and cloudy regions. It might be deceptive to just imagine the mean cloud thickness as the contribution to the redness of the spectrum, as this oversimplifies the situation by neglecting multi-dimensional effects compared to a purely 1D interpretation.

\subsection*{Additional results of the sensitivity tests}
\label{ch.s2}

This section presents sensitivity tests of our models with different cloud parameters and the strength of the bottom drag. 

\noindent \underline{\it Choice of cloud parameters}

\begin{figure}
    \centering
    \includegraphics[width=\linewidth]{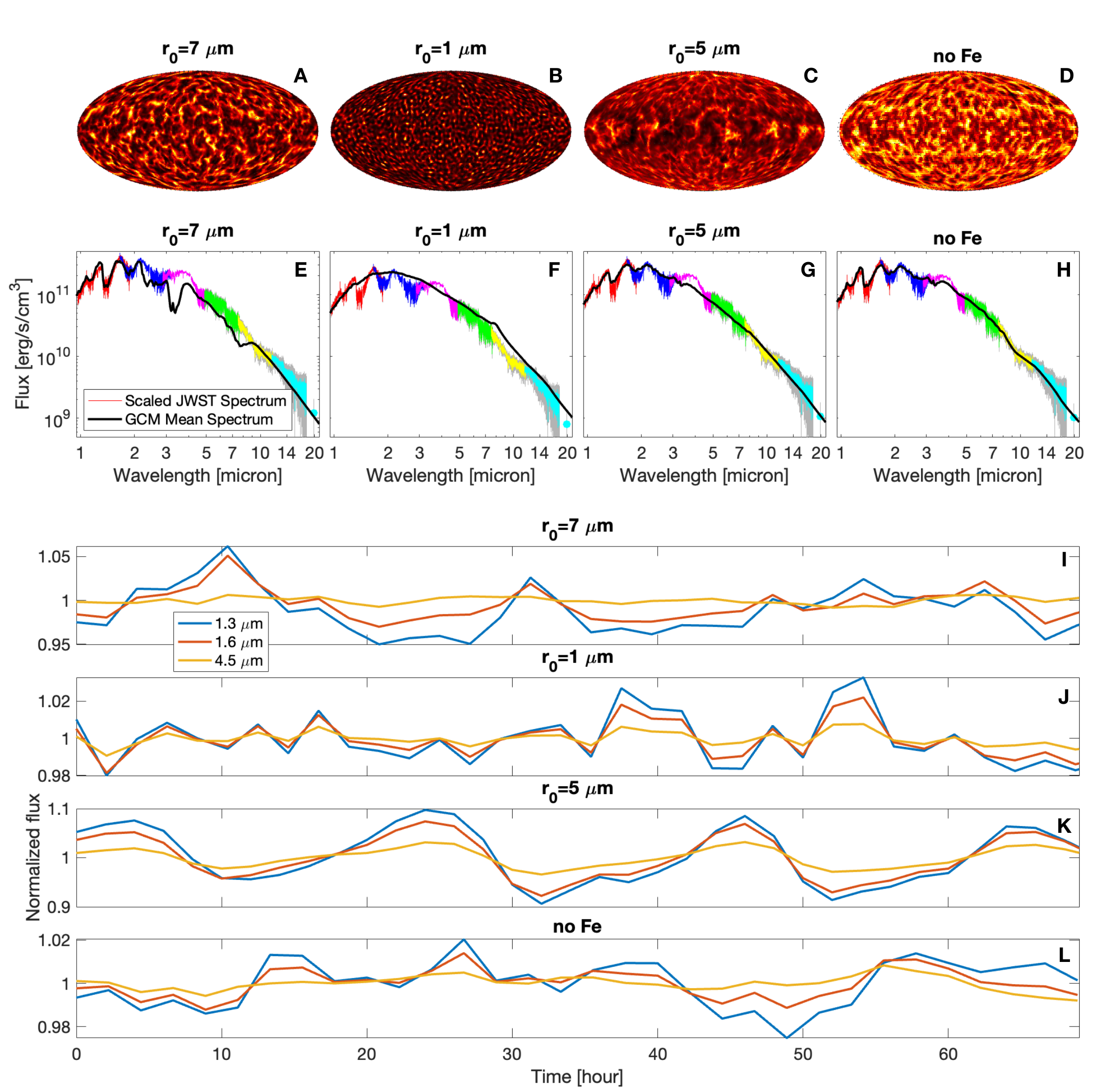}
    \caption{\add \textbf{Summary plots from GCM tests with one silicate mode of $r_0=7$, $1$, and $5~\mu$m, as well as for a test with the same silicate cloud setup as in the main text but excluding Fe clouds in the GCM.}  \textit{Panels A, B, C and D:} instantaneous global thermal flux maps for different tests. \textit{Panels E, F, G, and H:} Data-model comparison of the spectra. \textit{Panels I, J, K and L:} normalized spectroscopic light curves at a few selected wavelengths. 
    }\label{fig.summary_all}
\end{figure}

As discussed in the Materials and Methods section, we have performed models with a single size mode for both MgSiO$_3$ and Fe with varying $r_0$ and $\sigma$. The qualitative result is that smaller particle sizes tend to generate thicker clouds and redder spectra, because, given a roughly constant Si and Fe elementary budget, smaller particle size leads to more abundant number particle densities which exert stronger cloud radiative feedback; meanwhile, the more vigor circulation and the lower settling speeds naturally result in thicker clouds and redder spectra. Here we show two representative examples with $r_0=7~\mu$m, $\sigma=0.5$ and with $r_0=1~\mu$m, $\sigma=0.5$ in fig. \ref{fig.summary_all}. 

In the former case, the modeled spectrum qualitatively reproduces the red color at near-infrared up to about 2 $\mu$m but largely mismatches the observed spectrum at longer wavelengths. The mean atmosphere PT structure is cold such that too much CH$_4$ absorption appears near 3.3 and 7.8 $\mu$m. On the light curves, the short-term variability amplitudes are relatively small (for example, only 15\% at 1.3 $\mu$m) compared to results in the main text or the observed amplitude, and the light curve is much less regular over the rotational timescale. The thermal flux map of this model suggests that the cloud patchiness does not evolve on a global scale, limiting the variability amplitudes. In the latter case shown in fig. \ref{fig.summary_all}, the spectrum is far redder and has far weaker absorption features than the observed one. The variability amplitudes are even smaller and even more irregular and the light curves are more irregular over the rotational timescale, which can be immediately understood from the thermal map in which brightness patterns are even smaller. 

We qualitatively discuss the underlying dynamics of the above results and motivate the need to incorporate a small particle mode in the model presented in the main text. A rough timescale estimate for the tropical wave to fully develop is $L/c$, where $L$ is a relevant horizontal wave length scale, which in our situation could be taken as either the meridional width $\sqrt{\beta/c}$ or the zonal length that extends to the planetary radius, and $c$ is the horizontal wave phase speed. This timescale is on the order of $10^4$ to $10^5$ s for conditions relevant here. If other processes significantly disrupt the local cloud and temperature over a comparable or shorter timescale, the global-scale tropical wave cannot fully form and therefore the light-curve amplitudes are expected to be smaller. Such a process includes a local oscillation of temperature and cloud structure via the coupling between convection, cloud formation, and cloud radiative feedback as demonstrated in a time-dependent 1D model \cite{tan2019} and convection resolving model \cite{lefevre2022}. The column oscillation timescale is shorter when the particles are large and at relevant conditions it may reach a few hours \cite{tan2019}. The case with $r_0=7~\mu$m shown in fig. \ref{fig.summary_all} belongs to this category wherein column oscillation likely disrupts the global-scale wave formation and clouds form on a scale smaller than a planetary radius.

The case with small particles belongs to another regime where clouds are extremely well-mixed in the photosphere. As elucidated in \cite{tan2021bd1,tan2021bd2}, the horizontal differential heating that drives the large-scale wave from cloud radiative feedback requires clouds to be patchy--which requires clouds able to sink below the photosphere over a reasonable time. If clouds are so well-mixed that no significant horizontal differential heating can be achieved from the large-scale flow, the equatorial giant waves are suppressed. The spatial structure in the thermal map for $r_0=1~\mu$m of fig. \ref{fig.summary_all} illustrates the lack of large-scale structure, but only small-scale turbulence is retained in this regime, which corresponds to the extremely red spectrum and low-amplitude light curves. 

Using our initial model setup with a single mode for both MgSiO$_3$ and Fe, we systematically vary the particle size from $r_0=0.5$ to 10$~\mu$m and the results qualitatively agree with the above argument. The case with an intermediate particle size of $r_0=5~\mu$m generates observables with a closed match to the observed ones among these tests. However, the light-curve amplitudes reach only a maximum {\add of 20\% at 1.3$~\mu$m during its long-term evolution},  still far less than the observed 40\%. The synthetic spectrum has a good match in the near-infrared up to about 2 $\mu$m with a similar quality as in Fig. \ref{fig.summary}, but flux at longer wavelength was under-estimated.

These initial attempts triggered a search for a slightly more sophisticated cloud treatment that enhances the equatorial wave formation to better match the observed variability amplitude as well as a better explanation of the spectrum. Having some clouds settling faster but retaining the rest lofted in the atmosphere was a natural choice. This was motivated by cloud microphysics calculations that cloud undergoes nucleation, condensation growth, and coagulation, forming a broad size distribution \cite{gao2018,helling2019,gao2021}. The atmospheric transport should act individually on each particle size bin, rather than on the whole size distribution. This motivates the separation of one single mode into two different modes to represent this effect with a minimum complexity. To avoid further complications, Fe remains a single-size mode because it forms at deeper pressures than MgSiO$_3$ and has a smaller opacity impact. The larger mode of MgSiO$_3$ with $r_0=7~\mu$m in the model presented in the main text essentially depletes the majority of the cloud mass (which is set by the assumption that all Si elements in the solar composition condense) and remains a fraction of small particles ($r_0=0.7~\mu$m). These small particles have long settling timescales compared to the wave-formation timescale and just strong enough opacity feedback to favor wave formation, but their mixing ratio is not massive enough to result in too strong mixing (as in the case of $r_0=1~\mu$m in fig. \ref{fig.summary_all}) to suppress wave formation and completely redden the spectrum. In this framework, we have tested multiple choices of cloud sizes, and the case shown in the main text represents our best match to observations. Nevertheless, the model results with certain variations of the cloud parameters still agree with our qualitative argument, suggesting a robust dynamical mechanism. Finally, to reproduce the 10$~\mu$m silicate feature, a third tiny mode of MgSiO$_3$ is included. But this mode has negligible opacity at other wavelengths and is essentially passive to the dynamics and radiative flux. 

{\add Lastly, we perform a GCM model with only the silicate clouds of the same setup as the model presented in the main text, but excluding the Fe clouds. Results are summarized in panels D, H and L in fig. \ref{fig.summary_all}. The modeled low-resolution spectrum remains quantitatively similar to that of the standard GCM but the spatial cloud distribution does not exhibit the global-scale equatorial waves and is distinct from that of the standard GCM, as shown in the thermal flux map for the case of ``no Fe". As a result, the light curves have quite small amplitudes at different wavelengths, which does not agree with the observations. The lack of Fe opacity at slightly deeper atmospheres likely influences the wave formation processes. This test highlights the importance of considering Fe clouds in modeling the circulation, although Fe clouds might not be directly visible from spectra.  }

\noindent \underline{\it Varying the bottom drag strength}

\begin{figure}
    \centering
    \includegraphics[width=\linewidth]{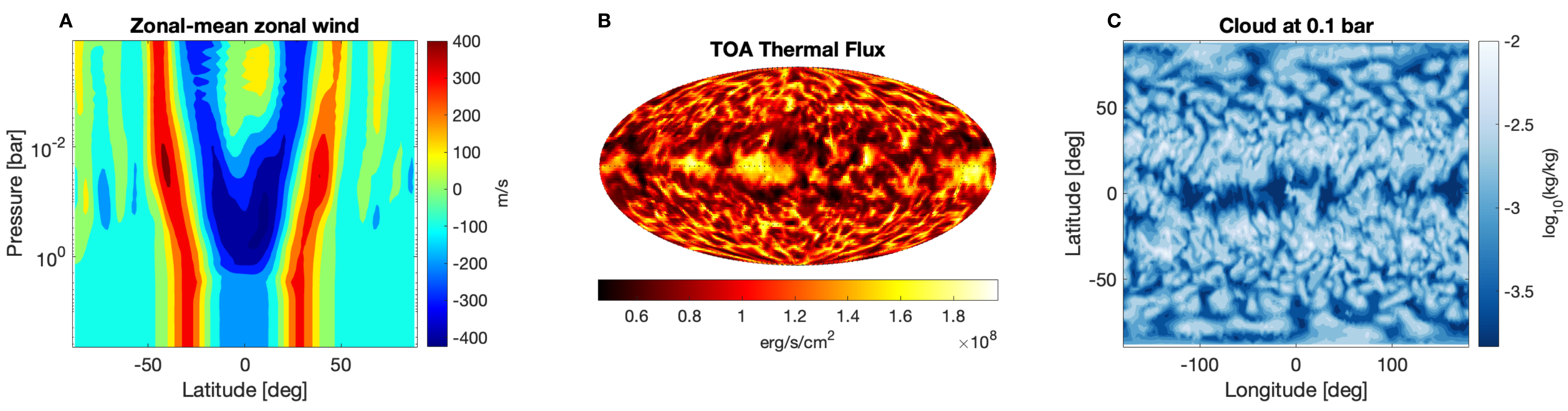}
    \caption{\textbf{Maps of the GCM outputs for the model similar to that presented in the main text, but with a longer bottom drag timescale $\tau_{\rm drag}=10^7$ s (weaker drag dissipation). } \textit{Panel A:} time-averaged zonal-mean zonal wind as a function of latitude and pressure, showing the jet structure. {\add \textit{Panel B:} the map of the instantaneous atmospheric thermal flux. \textit{Panel C:} total cloud mixing ratio on a logarithmic scale at 0.1 bar, at a simulated time corresponding to the thermal flux map. }
    }\label{fig.mapsd7}
\end{figure}

In conditions relevant to VHS 1256B, the interior is much hotter than that of Jupiter and we expect the interactions between the interior and the upper atmosphere may effectively suppress large-scale flow near the top of the convective zone, which is crudely parameterized as a bottom drag here \cite{schneider2009,showman2013bd}. Therefore, the main models have assumed a strong drag \cite{tan2022}. But we still test weaker bottom drag strengths by varying $\tau_{\rm drag}$ to be $10^6$ and $10^7$ s. The key results, including the formation of giant equatorial waves, relatively strong mixing of clouds, and the reddening of the spectrum  are qualitatively similar to those shown for the main model, as shown in fig. \ref{fig.mapsd7} for the case with $\tau_{\rm drag}=10^7$ s. The weak drag promotes the formation of deep zonal jets that extend to the lower boundary with speeds on the order of a few hundred ${\rm ms^{-1}}$. However, the jets cannot be visualized from the cloud and thermal flux maps, which instead are mostly shaped by the equatorial waves.

\begin{figure}
    \centering
    \includegraphics[width=0.8\linewidth]{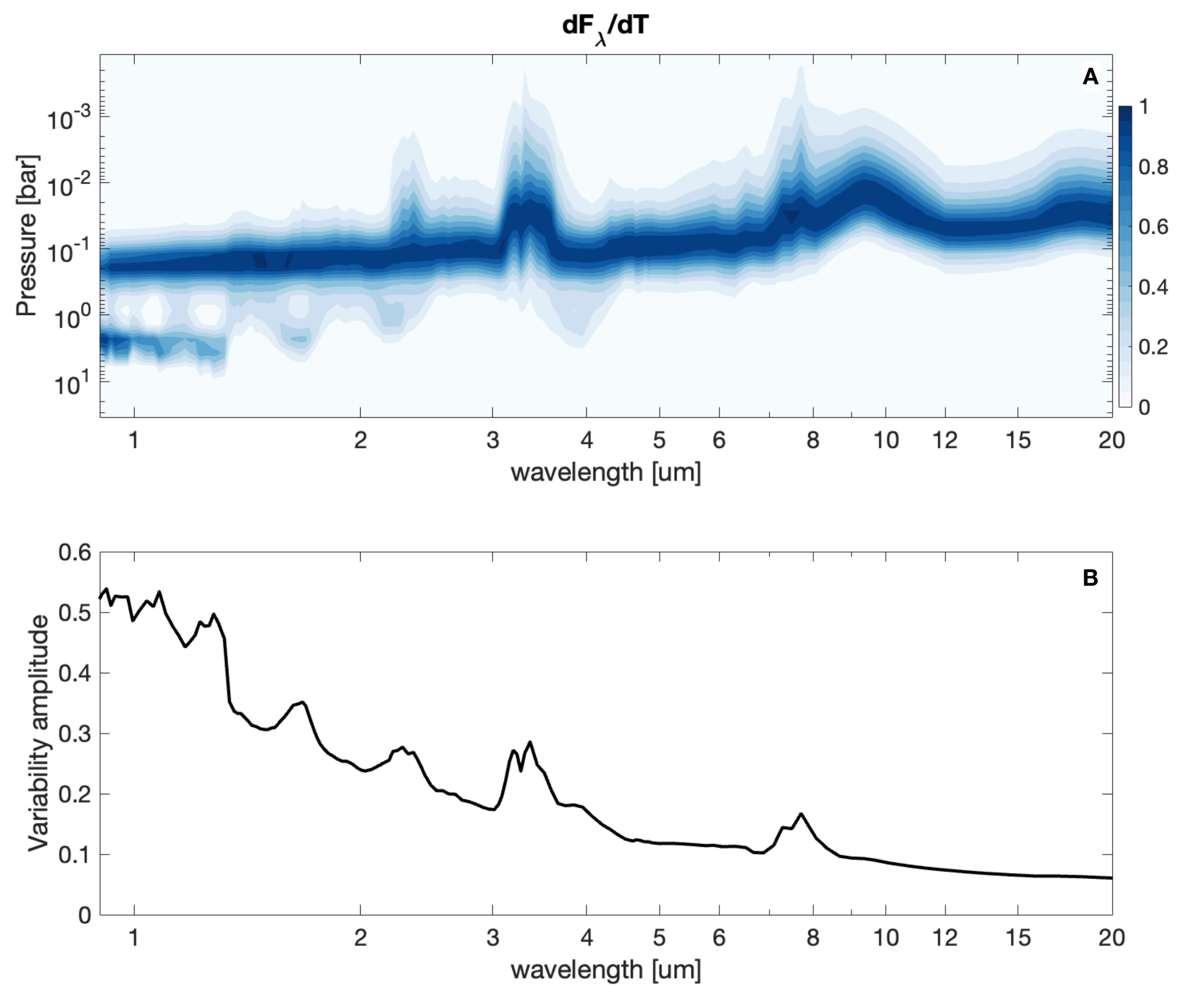}
    \caption{\textbf{Contribution function and maximum light curve deviations of the GCM.} Panel A: global-mean contribution function $dF_{\lambda}/dT$ as a function of wavelength and pressure from an instantaneous output of the GCM. Panel B: maximum deviation of the normalized light curves at Epoch 1.
    }\label{fig.CF_amp}
\end{figure}

\begin{figure}
    \centering
    \includegraphics[width=\linewidth]{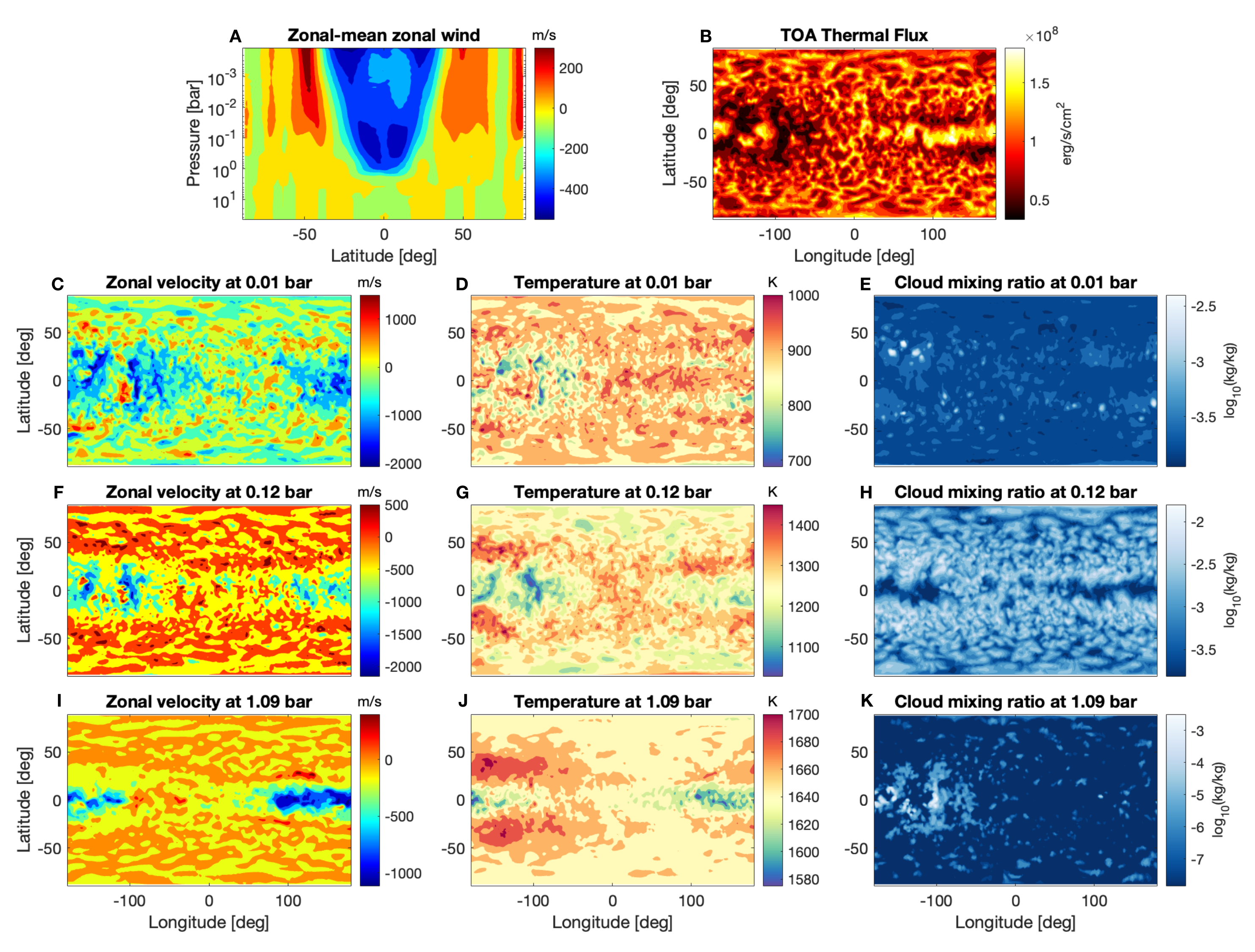}
    \caption{{\add \textbf{Maps of the GCM outputs for the model presented in the main text, taken at the simulated time of Epoch 1. } \textit{Panel A:} time-averaged zonal-mean zonal wind as a function of latitude and pressure, showing the jet structure. \textit{Panel B:} the map of the instantaneous top-of-atmosphere (TOA) thermal flux. \textit{Panels C to K:} horizontal maps of instantaneous zonal velocity fields (left column), temperature fields (middle column), and total cloud mixing ratio on a logarithmic scale (right column), taken at a simulated time corresponding to the thermal flux map. These maps are shown at pressure levels of 0.01, 0.12, and 1.09 bar (see titles above each panel).  These pressure levels are representative of the global-mean emission layers for a wide range of wavelengths, as shown in the contribution function of fig. \ref{fig.CF_amp}.}
    }\label{fig.maps}
\end{figure}

\clearpage 

\end{document}